# Odd-Parity Quasiparticle Interference in the Superconductive Surface State of $UTe_2$

Shuqiu Wang[1, 2§], Kuanysh Zhussupbekov[3,4§], Joseph P. Carroll[3,4§], Bin Hu[4], Xiaolong Liu[5], Emile Pangburn[6], Adeline Crepieux[7], Catherine Pepin[6], Christopher Broyles[8], Sheng Ran[8], Nicholas P. Butch[9], Shanta Saha[9,10], Johnpierre Paglione[9,10], Cristina Bena[6], J.C. Séamus Davis[1,3,4,11] and Qiangqiang Gu[4,12]

1. *Clarendon Laboratory, University of Oxford, Oxford, OX1 3PU, UK*
2. *H. H. Wills Physics Laboratory, University of Bristol, Bristol, BS8 1TL, UK*
3. *Department of Physics, University College Cork, Cork T12 R5C, IE*
4. *LASSP, Department of Physics, Cornell University, Ithaca, NY 14850, USA*
5. *Department of Physics, University of Notre Dame, Notre Dame, IN 46556, USA*
6. *Institut de Physique Théorique, Université Paris Saclay, CEA CNRS, Gif-sur-Yvette, France*
7. *Aix Marseille Univ, Université de Toulon, CNRS, CPT, Marseille, France*
8. *Department of Physics, Washington University. St Louis, MO 63130, USA*
9. *Maryland Quantum Materials Center, University of Maryland, College Park, MD 20742, USA*
10. *Canadian Institute for Advanced Research, Toronto, Ontario M5G 1Z8, CA*
11. *Max-Planck Institute for Chemical Physics of Solids, D-01187 Dresden, DE*
12. *Tsung-Dao Lee Institute, Shanghai Jiao Tong University, Shanghai, 200240, CN*

§ *These authors contributed equally to this project*

**Although no known material exhibits intrinsic topological superconductivity, wherein spin-triplet odd-parity electron pairing occurs, $UTe_2$ is now the leading representative of this class. Conventionally, the parity of the superconducting order parameter may be established by using Bogoliubov quasiparticle interference (QPI) imaging. However, odd-parity superconductors should support a topological quasiparticle surface band (QSB) at energies within the maximum superconducting energy gap. QPI would then be dominated by the electronic structure of the QSB and only reveal the characteristics of the bulk order parameter excursively. Here, we visualize quasiparticle interference patterns of $UTe_2$ and find that, at the (0-11) cleave surface, a new band of Bogoliubov quasiparticles appears only in the superconducting state. QPI visualization then allows study of dispersion of states within this QSB, which we demonstrate exists only within the range of Fermi momenta projected onto the (0-11) surface. Finally, we develop a theoretical framework to predict the QPI signatures of such a QSB at the (0-11) surface of $UTe_2$. Its predictions are most consistent with the**

**experimental results if the bulk superconducting gap function exhibits time-reversal conserving, odd-parity, *a*-axis nodal, $B_{3u}$ symmetry.**

The spin $-1/2$ electrons in superconductive materials can bind into a spin-zero singlet, or spin-one triplet[1,2] eigenstate. In the former case, the momentum $\mathbf{k}$ dependence of electron pairing potentials $\Delta(\mathbf{k})$ has even-parity, $\Delta(\mathbf{k}) = \Delta(-\mathbf{k})$, while in the latter its parity is odd, $\Delta(-\mathbf{k}) = -\Delta(\mathbf{k})$. Superfluid $^3He$[3] is the only material whose $\Delta(\mathbf{k})$ has definitely been identified as odd-parity, spin-triplet. If any such a superconductor exists, the electron pair potential is a matrix $\Delta_{\mathbf{k}} \equiv \begin{pmatrix} \Delta_{\mathbf{k}\uparrow\uparrow} & \Delta_{\mathbf{k}\uparrow\downarrow} \\ \Delta_{\mathbf{k}\downarrow\uparrow} & \Delta_{\mathbf{k}\downarrow\downarrow} \end{pmatrix}$ representing pairing with all three spin-1 eigenstates $(\uparrow\uparrow, \downarrow\downarrow, \uparrow\downarrow + \downarrow\uparrow)$, or equivalently $\Delta(\mathbf{k}) \equiv \Delta(\mathbf{d} \cdot \boldsymbol{\sigma}) i\sigma_2$ in $\mathbf{d}$-vector notation where $\sigma_i$ are Pauli matrices. $UTe_2$ is now widely surmised[4-6] to be such an odd-parity, spin-triplet intrinsic topological superconductor. Having $D_{2h}$ crystal symmetry and some degree of spin-orbit coupling, $UTe_2$ could, in theory, exhibit four possible odd-parity $\Delta(\mathbf{k})$ symmetries: $A_u$, $B_{1u}$, $B_{2u}$ and $B_{3u}$[6-8]. If extant, the $A_u$ phase would be fully gapped and preserve time reversal symmetry (akin to the B-phase of superfluid $^3He$[3]), while the $B_{1u}$, $B_{2u}$ and $B_{3u}$ phases also preserve time-reversal symmetry and would have point nodes in $\Delta(\mathbf{k})$ along the three orthogonal lattice axes provided a Fermi surface (FS) exists in these directions (akin to the hypothetical planar-phase of superfluid $^3He$[9]). Linear combinations of these four states might, when accidentally degenerate, break time-reversal symmetry generating distinct chiral $\Delta(\mathbf{k})$. For $UTe_2$, the challenge is to determine definitely which, if any, of these states exist.

Of course, it is the normal state electronic structure of $UTe_2$ that forms the basis upon which $\Delta(\mathbf{k})$ phenomenology emerges at lower temperatures. Atomic-resolution differential tunneling conductance $g(\mathbf{r}, V) \equiv dI/dV(\mathbf{r}, V)$ imaging visualizing the density-of-states $N(\mathbf{r}, E)$ and its Fourier transform $g(\mathbf{q}, E) \propto N(\mathbf{q}, E)$ can be used to establish those electronic-structure characteristics. A conventional model of the bulk first Brillouin zone (BZ) of $UTe_2$ sustaining a two-band FS as now widely hypothesized[6-8,10,11] is shown in Fig. 1a, while its contours at $k_z = 0$ are presented in Fig. 1b. Quantitative predictions for the normal state QPI in $UTe_2$ then require a Hamiltonian $H_{\mathrm{UTe}_2} = \begin{pmatrix} H_{U-U} & H_{U-Te} \\ H_{U-Te}^{+} & H_{Te-Te} \end{pmatrix}$ such that $H_{U-U}$ and $H_{Te-Te}$ describe respectively the two uranium and tellurium orbitals and $H_{U-Te}$ their hybridization (Methods). In Fig. 1b the intensity of each curve qualitatively represents the hybridized U 5$f$ orbital spectral weight in $k_z$ = 0 plane determined by quantum oscillations[12]. From this one might anticipate strong scattering interference with a sextet of wavevectors $\mathbf{p}_i: i = 1 - 6$, as indicated by the arrows:

| Wavevector | $\mathbf{p}_1$ | $\mathbf{p}_2$ | $\mathbf{p}_3$ | $\mathbf{p}_4$ | $\mathbf{p}_5$ | $\mathbf{p}_6$ |
|---|---|---|---|---|---|---|
| Coordinate $(\frac{2\pi}{a}, \frac{2\pi}{b})$ | (0.29,0) | (0.43, 1) | (0.29,2) | (0, 2) | (−0.14, 1) | (0.57,0) |

Table 1. Anticipated sextet of QPI wavevectors viewed in the (001) plane based on our band structure model.

However, the natural cleave surface of $UTe_2$ crystals is not (001) but rather[13] (0-11), here shown schematically in Fig. 1c. It is this surface that the scan-tip approaches perpendicularly. Its lattice vectors $\mathbf{a}_1$, $\mathbf{a}_2$ are identified in the top-right inset to Fig. 1d alongside the inter-tellurium chain distance $c^* = 0.76$ nm. The corresponding reciprocal lattice vectors $\mathbf{b}_1$, $\mathbf{b}_2$ are shown in the bottom-left inset to Fig. 1d, which is $T(\mathbf{q})$, the Fourier transform of $T(\mathbf{r})$. To clarify the normal state band structure and quasiparticle interference viewed from (0-11) plane, in Fig. 1e we first present the $\mathbf{k}$-space joint density of states $J(\mathbf{q}, E)$ calculated at the (001) plane using our $UTe_2$ FS that takes into account the uranium $f$ orbital spectral weight (Fig. 1b). The sextet of scattering wavevectors $\mathbf{p}_i: i = 1 - 6$ derived heuristically above are revealed as primary peaks in $J(\mathbf{q}, E)$**.** In Fig. 1f we present $J(\mathbf{q}, E)$ for the same band-structure model but viewed along the normal to the (0-11) plane (Methods and Extended Data Fig. 1). Here the $y$-coordinates of the (0-11) sextet become $\mathbf{q}_{1,y} = \mathbf{p}_{1,y} sin\theta$ where $\theta = 24°$and $c^*$ is the (0-11) surface $y$:$z$-axis lattice periodicity (colored arrows in Fig. 1f).

| Wavevector | $\mathbf{q}_1$ | $\mathbf{q}_2$ | $\mathbf{q}_3$ | $\mathbf{q}_4$ | $\mathbf{q}_5$ | $\mathbf{q}_6$ |
|---|---|---|---|---|---|---|
| Coordinate $(\frac{2\pi}{a}, \frac{2\pi}{c^*})$ | (0.29,0) | (0.43, 0.5) | (0.29,1) | (0, 1) | (−0.14, 0.5) | (0.57,0) |

Table 2. Anticipated sextet of QPI wavevectors viewed in the (0-11) plane based on our band structure model.

This QPI sextet $\mathbf{q}_i$ is quantitatively consistent with the precise $N(\mathbf{q}, E)$ and $J(\mathbf{q}, E)$ calculations presented below (Methods and Extended Data Figs. 5-7), and is pivotal to the remainder of our study.

Classically, odd-parity superconductors should exhibit zero-energy surface Andreev bound states[14-18] (SABS) which are generated by the universal π-phase-shift during Andreev reflections from the odd-parity $\Delta_\mathbf{k}$ (Methods). Hence, observation of a zero-energy SABS at an arbitrary crystal surface of a superconducting material would indicate that its $\Delta_\mathbf{k}$ has odd-parity. More intriguingly, intrinsic bulk topological superconductivity[19,20] exists most simply in the case of odd-parity spin-triplet superconductors. A definitive characteristic[21,22] of such intrinsic topological superconductor (ITS) would be a topological quasiparticle surface band (QSB) with momentum-energy relationship $\mathbf{k}(E)$ existing only for energies $|E| \leq \Delta$ within the maximum superconducting energy gap[21,23-34]. In $UTe_2$, there is now firm evidence from the pronounced zero-energy Andreev conductance[35] for the presence of a QSB at the (0-11) surface. Hence, QPI visualization studies and analyses for $UTe_2$ must take cognizance of the **k-**space structure of any such QSB.

In that context, we next consider Bogoliubov QPI imaging, a recognized technique for $\Delta(\mathbf{k})$ determination in complex superconductors[21, 36-43], in the superconducting state at temperatures much lower than the $UTe_2$ superconducting transition temperature. In this material, the $A_u$ state should be completely gapped on both Fermi surfaces while $B_{1u}$, $B_{2u}$ and $B_{3u}$ states could exhibit point nodes along the $k_z$ −axis, $k_y$ −axis and $k_x$ −axis, respectively. These bulk in-gap Bogoliubov eigenstates are described by the dispersion

$$E_\mathbf{k} = \sqrt{\xi_\mathbf{k}^2 + \Delta^2(|\mathbf{d}(\mathbf{k})|^2 \pm |\mathbf{d}(\mathbf{k}) \times \mathbf{d}^*(\mathbf{k})|)} \quad (1)$$

so that $\mathbf{k}$ -space locations of energy-gap zeros are defined in general by $|\mathbf{d}(\mathbf{k})|^2 \pm |\mathbf{d}(\mathbf{k}) \times \mathbf{d}^*(\mathbf{k})| = 0$. Formally, $A_u$ is fully gapped (nodeless). Modeling the pair potential magnitude $|\Delta_\mathbf{k}|$ for each order parameter throughout the $k_z = 0$ (001) BZ in Fig. 2a yields nodes at the dark blue regions where $|\Delta(\mathbf{k})|$ approaches 0. Thus, while $A_u$ supports no energy-gap nodes by definition and $B_{1u}$ exhibits no energy-gap nodes in this model, there are

numerous nodes in highly distinct **k**-space nodal locations for $B_{2u}$ and $B_{3u}$. Figure 2b presents a schematic of the bulk FS with energy-gap nodal locations for $B_{1u}$, $B_{2u}$ and $B_{3u}$ from Eqn. 3, shown as yellow dots.

Under these circumstances, to generate QPI predictions for the QSB in $UTe_2$ we use the Hamiltonian

$$H(\mathbf{k}) = \begin{pmatrix} H_{\mathrm{UTe_2}}(\mathbf{k})\otimes I_2 & \Delta(\mathbf{k})\otimes I_4 \\ \Delta^{+}(\mathbf{k})\otimes I_4 & -H^{*}_{\mathrm{UTe_2}}(-\mathbf{k})\otimes I_2 \end{pmatrix} \tag{2}$$

where the order parameter is $\Delta(\mathbf{k}) = \Delta_0(\mathbf{d}\cdot\boldsymbol{\sigma})i\sigma_2$ and $I_2$, $I_4$ are the unit matrices. We consider the order parameters: $A_u$, $B_{1u}$, $B_{2u}$, and $B_{3u}$ (Methods and Extended Data Fig. 1) but, because $A_u$ and $B_{1u}$ are non-nodal, here we focus primarily on $B_{2u}$ and $B_{3u}$:

$$\mathbf{d}_{B_{2u}} = \big(C_1 sin(k_z c), C_0 sin(k_x a) sin(k_y b) sin(k_z c), C_3 sin(k_x a)\big) \tag{3a}$$

$$\mathbf{d}_{B_{3u}} = \big(C_0 sin(k_x a) sin(k_y b) sin(k_z c), C_2 sin(k_z c), C_3 sin(k_y b)\big) \tag{3b}$$

where *a, b, c* are lattice constants, and $C_0 = 0$, $C_1 = 300$ μeV, $C_2 = 300$ μeV, and $C_3 = 300$ μeV. The unperturbed bulk Green's function is then: $G_0(\mathbf{k}, E) = [(E + i\eta)I - H(\mathbf{k})]^{-1}$ ($\eta$ = 100 μeV) with the corresponding unperturbed spectral function: $A_0(\mathbf{k}, E) = -1/\pi \, \mathrm{Im}\, G_0(\mathbf{k}, E)$. While obtaining the $G_0(\mathbf{k}, E)$ is straightforward, calculating the (0-11) surface Green's functions $G_S(\mathbf{k}, E)$ and spectral functions $A_S(\mathbf{k}, E)$ is significantly more difficult. The surface Green's function characterizes a semi-infinite system with broken translation symmetry and therefore cannot be calculated directly. Here we use a technique in which we model the surface using a strong planar impurity[44-46]. In the limit of an infinite impurity potential, the impurity plane splits the system into two semi-infinite spaces. Then only wavevectors in the (0-11) plane remain good quantum numbers. The effect of the planar-impurity can then be exactly calculated using the T-matrix formalism which gives one access to the surface Green's function of the semi-infinite system. Details of this procedure can be found in Methods and Extended Data Fig. 2. The predicted surface quasiparticle spectral function, $A_S(\mathbf{k}, E)$, calculated using the above method for the $B_{1u}$, $B_{2u}$ and $B_{3u}$ order parameters also appear in Methods and Extended Data Fig. 3. For Bogoliubov QPI predictions at the (0-11) surface of $UTe_2$, we use a localized impurity potential $\hat{V} = V\tau_z \otimes I_8$ where $V = 0.2$ eV (Methods and Extended Data Fig. 6), and determine the exact solution for the perturbed generalized surface Green's function $g_S(\mathbf{q},\mathbf{k},E)$ using the T-matrix $T(E)=\left(I-\hat{V}\int\frac{d^2\mathbf{k}}{S_{BZ}}G_s(\mathbf{k},E)\right)^{-1}\hat{V}$. Then the QPI patterns for the $UTe_2$ QSB are predicted directly using:

$$N(\mathbf{q},E)=\frac{i}{2\pi}\int\frac{d^2\boldsymbol{k}}{S_{BZ}}Tr[g_S(\mathbf{q},\mathbf{k},E)] \tag{4}$$

where

$$g_S(\mathbf{q},\mathbf{k},E)=G_s(\mathbf{q},E)T(E)G_s(\mathbf{q}-\mathbf{k},E)-G_s^*(\mathbf{q}-\mathbf{k},E)T^*(E)G_s^*(\mathbf{q},E) \tag{5}$$

By calculating the trace over particle-hole space on $g_S(\mathbf{q},\mathbf{k},E)$, the obtained $N(\mathbf{q},E)$ is in general a complex quantity, all simulations presented herein are therefore $|N(\mathbf{q},E)|$. The predicted QSB spectral function, $A_S(\mathbf{k},E)$, joint density of states $J(\mathbf{q},0)$, and density of states spectra for a $B_{2u}$-QSB and $B_{3u}$-QSB within the (0-11) surface Brillouin zone (SBZ), appear in Methods and Extended Data Fig. 5. We further take into account the $\mathbf{q}$-space sensitivity of our scan tip by applying a 2D Gaussian filter to the $N(\mathbf{q},E)$ calculated using Eq. (4) (Methods and Extended Data Fig. 6). Additionally, we also discuss alternative, symmetry-allowed, gap structure models and derive their resulting $A_0(\mathbf{k},E)$, $A_S(\mathbf{k},E)$, and $J(\mathbf{q},0)$ (Methods and Extended Data Fig. 7), finding them indistinguishable from the results presented in Extended Data Fig. 5. Ultimately, the existence of these specific QPI characteristics in $UTe_2$ would provide strong confirmation of both a superconductive QSB and its foundational odd-parity bulk order-parameter.

Experimental exploration of such phenomena is challenging in $UTe_2$ and several key technical advances were employed to improve on previous studies[35]. First we identified regions where the QPI signal predominantly originates from a single type of identical impurity ($Te_2$ vacancies); second, the field of view (FOV) studied here is larger thus improving the $\mathbf{q}$-space resolution; third by using Andreev tunneling the energy resolution is ~10 μeV [13] and the QPI signal-to-noise ratio is strongly enhanced (see below). Figure 3a then shows a typical 66 nm square field-of-view (FOV) topography of the (0-11) cleave surface which can be studied both in the normal and superconducting states. Figure 3b shows characteristic $dI/dV$ spectra measured with a superconductive tip in both the normal state at 4.2 K and the superconducting state at 280 mK, far below $T_C$. In the latter case, two intense joint-coherence peaks are located at $E = \Delta_{\mathrm{Nb}} + \Delta_{\mathrm{UTe_2}}$. More importantly, a high density of QSB quasiparticles allows efficient creation/annihilation of Cooper pairs in both superconductors, thus generating intense Andreev differential conductance[35] $a(\mathbf{r}, V) \equiv dI/dV|_A(\mathbf{r}, V)$ for $|V| < \Delta_{\mathrm{UTe_2}}/\mathrm{e} \sim 300\ \mu\mathrm{V}$ as indicated by blue vertical dashed lines (Methods and Extended Data Fig. 8). Compared to conventional normal-insulating-superconducting (NIS) tunneling using a normal metallic tip (Methods and Extended Data Fig. 9), this Andreev conductance provides a significant improvement in the energy resolution ($\delta E \sim 10$ μeV) of QSB scattering interference measurements. Comparing measured $g(\mathbf{r}, V)$:$g(\mathbf{q}, V)$ recorded in the normal state at 4.2 K (Fig. 3c) with measured $a(\mathbf{r}, V)$: $a(\mathbf{q}, V)$ in the superconducting state at 280 mK (Fig. 3d), both with identical FOV and junction characteristics, allows determination of which phenomena at the (0-11) surface emerge only due to superconductivity. Some peaks of the sextet are present in the normal state $g(\mathbf{q}, V)$ in Fig. 3c as they originate from scattering of the normal state band structure (Fig. 1b). The experimentally obtained normal state QPI differs from the $J(q, 0)$ calculations in Fig. 1f, as the former depends on spin and orbital selection rules while the latter is dependent only on the geometry of the bulk band structure. Instead, the complete predicted QPI sextet $\mathbf{q}_i$: $i = 1 - 6$ are only detected in the superconducting state and appears to rely on scattering between QSB states. The sextet wavevectors are highlighted by colored arrows in Fig. 3d. The experimental maxima in $a(\mathbf{q}, V)$ and the theoretically predicted $\mathbf{q}_i$ from Fig. 1f, are in excellent quantitative agreement with a maximum 3% difference between all their wavevectors. This demonstrates that the FS which dominates the bulk electronic structure of $UTe_2$ is also what controls QSB **k**-space geometry at its cleave surface. Furthermore, Fig. 3e reveals how the amplitudes of the superconducting state QPI are enhanced compared to the normal state measurements. The predominant effects of bulk superconductivity are the strongly enhanced arc-like scattering intensity connecting $\mathbf{q} = 0$ and $\mathbf{q}_5$, and the unique appearance of purely superconductive QPI at wavevector $\mathbf{q}_1$.

To visualize the QSB dispersion $\mathbf{k}(E)$ of $UTe_2$ we next use superconductive-tip $a(\mathbf{r}, V)$: $a(\mathbf{q}, V)$ measurements to image energy resolved QPI at the (0-11) cleave surface. Figures 4a-f presents the measured $a(\mathbf{r}, V)$ at $V$ =0 μV, 50 μV, 100 μV, 150 μV, 200 μV, 250 μV recorded at $T$ = 280 mK in the identical FOV as Fig. 3a. These data are highly typical of such experiments in $UTe_2$. Figure 4h contains the consequent scattering interference patterns $a(\mathbf{q}, V)$ at $V = 0$ μV, 50 μV, 100 μV, 150 μV, 200 μV, 250 μV as derived by Fourier analysis of Figs. 4a-f. Here the energy evolution of scattering interference of the QSB states is manifest. For comparison with theory, detailed predicted characteristics of $N(\mathbf{q}, E)$ for a $B_{2u}$-QSB and $B_{3u}$-QSB at the (0-11) SBZ are presented in Figs. 4g,i; here again energies range $E = 0$ μeV, 50 μeV, 100 μeV, 150 μeV, 200 μeV, 250 μeV. Each QPI wavevector is determined by maxima in the $N(\mathbf{q}, E)$ QPI pattern (colored circles in Figs. 4g-i); these phenomena are highly repeatable in multiple independent experiments (Methods and Extended Data Fig. 10). The general correspondence of $B_{3u}$-QSB theory to the experimental QPI data is striking. Significantly, the strongly enhanced QPI features occurring along the arc connecting $\mathbf{q} = 0$ and $\mathbf{q}_5$ (Fig. 4h) are characteristic of the theory for a $B_{3u}$-QSB (Fig. 4i). The arc connecting the $\mathbf{q}_1$ and $\mathbf{q}_2$ (Fig. 4i) is the consequence of projected FS scattering and it is irrelevant to the superconducting order parameter $\Delta(\mathbf{k})$. Most critically, however, the intense QPI appearing at wavevector $\mathbf{q}_1$ (red circles in Figs. 4h, i) is a characteristic of the $B_{3u}$ superconducting state, deriving from its geometrically unique nodal structure (Extended Data Fig. 5). Further analysis involving the calculation of the spin-resolved surface spectral function (Extended Data Fig. 4) establish that scattering at $\mathbf{q}_1$ is suppressed for $B_{2u}$ gap symmetry due to proscribed spin-flip scattering processes but is uniquely enhanced for $B_{3u}$ gap symmetry. Moreover, the appearance of scattering interference of QSB quasiparticles at $\mathbf{q}_1$ in the superconducting state (Fig. 3d) is as anticipated by theory[19,31] due to projection of $B_{3u}$ energy-gap nodes on the bulk FS (Fig. 2) onto the (0-11) crystal surface 2D Brillouin zone.

While the superconductive quasiparticle surface band of $UTe_2$ has now been rendered directly accessible to visualization (Figs. 3, 4), its precise topological categorization[19,20,23-34] depends on details of the normal state FS which have not yet been determined conclusively[10,11]. Nevertheless, major advances in empirical knowledge of both the QSB and the bulk $\Delta_{\mathbf{k}}$ symmetry of the putative topological superconductor $UTe_2$ have been achieved. By introducing superconductive scan-tip Andreev tunneling spectroscopy, which is specifically sensitive to the QSB of intrinsic topological superconductors, we visualize dispersive QSB scattering interference of $UTe_2$ (Figs. 3d, 4h). This reveals exceptional in-gap QPI patterns exhibiting a characteristic sextet of wavevectors $\mathbf{q}_i: i = 1 - 6$ (Figs. 3d,e) that we demonstrate are due to projection of the bulk superconductive band structure (Figs. 1a,b), mathematically equivalent to a rotation making the point-of-view perpendicular to the (0-11) plane (Fig. 1f). Thence, we find that, while $\mathbf{q}_2$ and $\mathbf{q}_6$ are weakly observable in the normal state and $\mathbf{q}_4$ is a Bragg peak of the (0-11) surface, features at $\mathbf{q}_5$ and $\mathbf{q}_6$ become strongly enhanced for superconducting state QPI at $|E| < \Delta$ (Fig. 3e). Most critically, intense QPI appears at wavevector $\mathbf{q}_1$ uniquely in the superconducting state (Figs. 3e, 4h). This complete QSB phenomenology (Figs. 3d,e; 4h) is, by correspondence with theory (Figs. 1; 2; 4g, i), most consistent with a $B_{3u}$-symmetry superconducting order-parameter. Collectively we identify the $B_{3u}$ state in particular: first, because its unique nodal structure enhances the spectral weight of the QSB responsible for the arc-like feature connected to $\mathbf{q}_5$ in the superconducting state (Fig. 4h) and, second, because $B_{3u}$ is the only state that produces intense QPI at wavevector $\mathbf{q}_1$ uniquely in the superconducting state (red circle in Figs. 4h, i).

These considerations indicate that $UTe_2$ sustains a 3D, odd-parity, spin-triplet, time-reversal-symmetry conserving, **a**-axis nodal superconducting order parameter (Fig. 2). Moreover we establish how this 3D $\Delta_{\mathbf{k}}$ on its host FS is projected onto the 2D SBZ, generating a superconductive in-gap QSB (Fig. 3d) consistent with general theory for intrinsic topological superconductors[21,22] and other related results[47]. Overall, the data indicate that the superconductive quasiparticle surface band QPI phenomenology (Fig. 4) is a direct consequence of the **k**-space geometry of the FS projected onto the crystal surface of $UTe_2$, reveal the existence and energy dispersion $\mathbf{k}_\sigma(E)$ of this exceptional in-gap QSB, and provide prefatorial evidence that its quasiparticle scattering interference is due to $B_{3u}$-symmetry bulk superconductivity in $UTe_2$. Most generally, the techniques initiated here represent a particularly promising new approach for the identification of intrinsic topological superconductors.

**Acknowledgements:** We are extremely grateful to D.-H. Lee for critical advice and guidance on superconductive topological surface band physics. We thank W. A. Atkinson, A. Carrington, P. J. Hirschfeld, and S. Sondhi for helpful discussions. Research at the University of Maryland was supported by the Department of Energy Award No. DE-SC-0019154 (sample characterization), the Gordon and Betty Moore Foundation's EPiQS Initiative through Grant No. GBMF9071 (materials synthesis), NIST, and the Maryland Quantum Materials Center. The work at Washington University is supported by McDonnell International Scholars Academy and the U.S. National Science Foundation (NSF) Division of Materials Research Award DMR-2236528. A.C. thanks the CALMIP supercomputing center for the allocation of HPC numerical resources through project M23023. S.W. acknowledges support from the Engineering and Physical Sciences Research Council (EPSRC) under Award EP/Z53660X/1. J.C.S.D. acknowledges support from the Royal Society under Award R64897. J.P.C., K.Z. and J.C.S.D. acknowledge support from Science Foundation Ireland under Award SFI 17/RP/5445. S.W. and J.C.S.D. acknowledge support from the European Research Council (ERC) under Award DLV-788932. Q.G., K.Z., J.P.C., S.W., B.H., and J.C.S.D. acknowledge support from the Moore Foundation's EPiQS Initiative through Grant GBMF9457.

**Author Contributions:** J.C.S.D., C.P. and C. Bena conceived and supervised the project. C. Broyles, S.R., N.P.B. S.S. and J.P. developed, synthesized, and characterized materials; C. Bena, A.C., E.P. and C.P. provided theoretical analysis of QSB electronic structure and QPI; K.Z., B.H., J.P.C., S.W., X.L. and Q.G. carried out the experiments; J.P.C., K.Z., S.W. and Q.G. developed and implemented analysis. J.C.S.D., C.P. and C. Bena wrote the paper with key contributions from S.W., K.Z., J.P.C., A.C., E.P. and Q.G. The paper reflects contributions and ideas of all authors.

**Competing interests:** The authors declare no competing interests.

## Figure captions

**FIG. 1 Fermi Surface and QPI Predictions for $UTe_2$: Projection to (0-11) surface.**

- **a.** Bulk Fermi surface (FS) of $UTe_2$ based on recent band structure models (Methods).
- **b.** Bulk $UTe_2$ FS intersecting the $k_z = 0$ plane. Highlighted with colored arrows are a sextet of scattering interference wavevectors $\mathbf{p}_i$, $i$ = 1-6 connecting spectral weight maxima in **k**-space derived heuristically from *f*-electron orbital contributions.
- **c.** Schematic of $UTe_2$ (0-11) cleave surface, whose normal is oriented to the crystal **b**-axis at $\theta \cong 24°$. Uranium (red) and two inequivalent tellurium atom sites (dark and light blue) overlaid on a *T*(**r**) image, revealing the tellurium chains of the (0-11) cleave surface.
- **d.** Typical topographic image *T*(**r**) of the (0-11) cleave surface of $UTe_2$. Top-right inset shows both the *x*-axis unit cell distance *a*, and the *y:z*-axis lattice periodicity *c**, as well as the (0-11) termination surface primitive lattice vectors, $\mathbf{a}_1$ and $\mathbf{a}_2$. Bottom-left inset, *T*(**q**), Fourier transform of *T*(**r**), shows the (0-11) reciprocal unit cell.
- **e.** Joint density of states ($J(\mathbf{q}, E)$) calculated using the model featured in **b** for $k_z = 0$ of the crystal termination layer (001). The sextet of scattering interference wavevectors $\mathbf{p}_i$, $i$ = 1-6 connecting maxima in **b** are overlaid.
- **f.** $J(\mathbf{q}, E)$ predicted for the (0-11) termination from the FS model of **b**. Rotation to the (0-11) plane corresponds to a change in *y*-axis coordinates $\mathbf{q}_{1,y} = \mathbf{p}_{1,y} sin\theta$. Here the sextet of QPI wavevectors $\mathbf{q}_i$, $i$ = 1-6, now viewed along the normal to (0-11), are overlaid.

**FIG. 2 Simple Models for $UTe_2$ $\Delta_\mathbf{k}$.**

- **a.** Magnitude of the $UTe_2$ superconductive energy-gap $|\Delta_\mathbf{k}|$ at $k_z = 0$ for the $B_{1u}$, $B_{2u}$, and $B_{3u}$ order parameters on the FS shown in Figs. 1a, b. The nodal locations occur within the dark blue regions where $|\Delta_\mathbf{k}| \rightarrow 0$. Note that $B_{1u}$ does not exhibit gap nodes in this model because the FS is open along the $k_z$-axis.

**b.** From ***a*** the theoretically predicted nodal locations for the $B_{1u}$, $B_{2u}$, and $B_{3u}$ order parameters on the FS shown in Figs. 1a, b are indicated by yellow dots. For $B_{1u}$, $\mathbf{d} \propto (\sin k_y b, \sin k_x a, 0)$ and zeros occur at $k_y = 0, \pm\frac{\pi}{b}$, $k_x = 0, \pm\frac{\pi}{a}$; for $B_{2u}$, $\mathbf{d} \propto (\sin k_z c, 0, \sin k_x a)$ and zeros occur at $k_z = 0, \pm\frac{\pi}{c}, k_x = 0, \pm\frac{\pi}{a}$; and for $B_{3u}$, $\mathbf{d} \propto (0, \sin k_z c, \sin k_y b)$ and zeros occur at $k_z = 0, \pm\frac{\pi}{c}$, $k_y = 0, \pm\frac{\pi}{b}$ (Methods and Extended Data Fig. 1). Alternative gap functions and consequent nodal locations are discussed in Methods and Extended Data Fig. 7.

**FIG. 3 QPI Visualization of $UTe_2$ Superconductive Quasiparticle Surface Band.**

**a.** Typical topographic image $T(\mathbf{r})$ of the (0-11) cleave surface of $UTe_2$.

**b.** Measured differential conductance in the $UTe_2$ normal state $g(V)$ at $T$=4.2 K; and Andreev differential conductance in the superconducting state $a(V)$ at $T$=280 mK. Intense Andreev conductance is observed at $V$ = 0.

**c.** Measured $g(\mathbf{r}, 0)$ and $g(\mathbf{q}, 0)$ at $T$=4.2 K in the $UTe_2$ normal state in the identical FOV as **a**. The setpoint is $V_s$ = 3 mV and $I$ = 200 pA.

**d.** Measured $a(\mathbf{r}, 0)$ and $a(\mathbf{q}, 0)$ at $T$=280 mK in the $UTe_2$ superconducting state in the identical FOV as **a** and **c**. Here a sextet of scattering interference wavevectors $\mathbf{q}_i$, $i$ = 1-6 from theoretical predictions are overlaid. The excellent correspondence between the predictions and the measured QPI data is striking, with all theory and experiment wavevectors $\mathbf{q}_1$, $\mathbf{q}_2$, $\mathbf{q}_3$, $\mathbf{q}_4$, $\mathbf{q}_5$ and $\mathbf{q}_6$ being within 3% of each other. This experimental detection of the sextet has been repeated multiple times (Methods and Extended Data Fig. 10). The set point is $V_s$ = 3 mV and $I$ = 200 pA.

**e.** Relative amplitudes of the sextet wavevectors in the normal and superconducting states. Comparison of $g(\mathbf{q}, 0)$ linecuts at $T$=4.2 K and $a(\mathbf{q}, 0)$ linecuts measured $T$=280 mK. The linecuts are taken horizontally in the **q** space indicated by white arrow in **d**. The linecuts have been normalized by their background intensities at 280 mK and 4.2 K. The intensities of $\mathbf{q}_5$ and $\mathbf{q}_6$ have been significantly enhanced in the superconducting state. Most importantly, $\mathbf{q}_1$ only appears in the superconducting state.

**FIG. 4 Quasiparticle Surface Band QPI for Δ(k) Identification in $UTe_2$.**

**a-f**. Measured *a*(**r**, *V*) at the (0-11) cleave plane of $UTe_2$ at bias voltages $|V| = 0\ \mu V, 50\ \mu V, 100\ \mu V, 150\ \mu V, 200\ \mu V, 250\ \mu V$. The setpoint is $V_s$ = 3 mV and *I* = 200 pA.

**g.** Predicted QPI patterns for a $B_{2u}$-QSB at the (0-11) SBZ of $UTe_2$ at energies $|E| = 0\ \mu eV, 50\ \mu eV, 100\ \mu eV, 150\ \mu eV, 200\ \mu eV, 250\ \mu eV$ (Methods and Extended Data Figs. 3-6). We take into account the finite radius of the scan tip in simulations by applying a 2D Gaussian to the $N(\mathbf{q}, E)$ maps (Methods and Extended Data Fig. 6). The existing QPI wavevector $\mathbf{q}_2$ is identified as the maxima position (brown circle) in the QPI simulation.

**h.** Measured *a*(**q**, *V*) at the (0-11) cleave plane of $UTe_2$ at bias voltages $|V| = 0\ \mu V, 50\ \mu V, 100\ \mu V, 150\ \mu V, 200\ \mu V, 250\ \mu V$. The setpoint is $V_s$ = 3 mV and *I* = 200 pA. These QPI data are derived by Fourier transformation of *a*(**r**, *V*) data in Figs. 4a-f. Each QPI wavevector in this FOV, $\mathbf{q}_1$ (red), $\mathbf{q}_2$ (brown) and $\mathbf{q}_5$ (cyan), is identified as the maxima position (colored circles) in the experimental QPI data. Particularly $\mathbf{q}_1$ is a characteristic only of the $B_{3u}$ superconducting state and it only exists inside the energy gap. $\mathbf{q}_1$ cannot be due to a pair density wave (Methods).

**i.** Predicted QPI patterns for a $B_{3u}$-QSB at the (0-11) SBZ of $UTe_2$ at energies $|E| = 0\ \mu eV, 50\ \mu eV, 100\ \mu eV, 150\ \mu eV, 200\ \mu eV, 250\ \mu eV$ (Methods and Extended Data Figs. 3-6). Each QPI wavevector, $\mathbf{q}_1$, $\mathbf{q}_2$ and $\mathbf{q}_5$, is identified as the maxima position (colored circles) in the QPI simulation. We take into account the finite radius of the scan tip in simulations by applying a 2D Gaussian to the $N(\mathbf{q}, E)$ maps (Methods and Extended Data Fig. 6).

## References

1 P. W. Anderson, P. Morel, Generalized Bardeen-Cooper-Schrieffer States, *Phys. Rev.* **123**, 1911 (1961).

2 R. Balian, N. R. Werthamer, Superconductivity with Pairs in a Relative *p*-Wave, *Phys. Rev.* **131**, 1553 (1963).

3 D. Vollhardt and P. Woelfle, *The Superfluid Phases of Helium 3* (Taylor & Francis, 1990).

4 D. Aoki et al., Unconventional Superconductivity in Heavy Fermion $UTe_2$, *J. Phys. Soc. Jpn.* **88**, 043702 (2019).

5 S. Ran et al., Nearly ferromagnetic spin-triplet superconductivity, *Science* **365**, 684-687 (2019).

6 D. Aoki et al., Unconventional superconductivity in $UTe_2$, *J. Phys.: Condens. Matter* **34**, 243002 (2022).

7 T. Shishidou et al., Topological band and superconductivity in $UTe_2$, *Phys. Rev. B* **103**, 104504 (2021).

8 J. Tei et al., Possible realization of quasiparticle crystalline superconductivity with time-reversal symmetry in $UTe_2$, *Phys. Rev. B* **107**, 144517 (2023).

9 G. Volovik, From elasticity tetrads to rectangular vielbein, *Annals of Physics* **447**, 168998 (2022).

10 A. G. Eaton et al., Quasi-2D Fermi surface in the anomalous superconductor $UTe_2$. *Nat. Commun.* **15**, 223 (2024).

11 C. Broyles et al., Revealing a 3D Fermi Surface Pocket and Electron-Hole Tunneling in $UTe_2$ with Quantum Oscillations, *Phys. Rev. Lett.* **131**, 036501 (2023)

12 T. I. Weinberger et al., Pressure-enhanced *f*-electron orbital weighting in $UTe_2$ mapped by quantum interferometry, *arXiv*. 2403.03946 (2024)

13 Q. Gu, J. P. Carroll, S. Wang, S. Ran, C. Broyles, H. Siddiquee, N. P. Butch, S. R. Saha, J. Paglione, J. C. Davis, X. Liu, Detection of a pair density wave state in $UTe_2$, *Nature* **618**, 921–927 (2023).

14 L.J. Buchholtz, G. Zwicknagl, Identification of *p*-wave superconductors, *Phys. Rev. B* **23**, 5788 (1981)

15 J. Hara, K. Nagai, A Polar State in a Slab as a Soluble Model of *p*-Wave Fermi Superfluid in Finite Geometry, *Prog. Theor. Phys.* **76**, 1237 (1986)

16 K. Honerkamp, M. Sigrist, Andreev Reflection in Unitary and Non-Unitary Triplet States, *J. Low Temp. Phys.* **111**, 895–915 (1998)

17 S. Kashiwaya, Y. Tanaka, Tunnelling effects on surface bound states in unconventional superconductors, *Rep. Prog. Phys.* **63**, 1641 (2000)

18 J. Sauls, Andreev bound states and their signatures, *Phil. Trans. R. Soc. A.* **376**, 20180140 (2018)

19 A. P. Schnyder et al., Classification of quasiparticle insulators and superconductors in three spatial dimensions, *Phys. Rev. B* **78**, 195125 (2008).

20 X.L. Qi and S.C. Zhang, Topological insulators and superconductors, *Rev. Mod. Phys*. **83**, 1057-1110 (2011).

21 J.S. Hofmann, R. Queiroz, A.P. Schnyder, Theory of quasiparticle scattering interference on the surface of topological superconductors, *Phys. Rev. B* **88**, 134505 (2013).

22 Y. Tanaka et al., Theory of Majorana Zero Modes in Unconventional Superconductors, *Prog. Theor. Exp. Phys*. **8**, 08C105 (2024).

23 M. Stone and R. Roy, Edge modes, edge currents, and gauge invariance in $p_x+ip_y$ superfluids and superconductors, *Phys. Rev. B* **69**, 184511 (2004).

24 S. B. Chung and S.-C. Zhang, Detecting the Majorana Fermion Surface State of $^3$He−*B* through Spin Relaxation, *Phys. Rev. Lett*. **103**, 235301 (2009).

25 Y. Tsutsumi, M. Ichioka, and K. Machida, Majorana surface states of superfluid $^3$He A and B phases in a slab, *Phys. Rev. B* **83**, 094510 (2011).

26 T. H. Hsieh and L. Fu, Majorana Fermions and Exotic Surface Andreev Bound States in Topological Superconductors: Application to $Cu_xBi_2Se_3$, *Phys. Rev. Lett.* **108**, 107005 (2012).

27 F. Wang and D.H. Lee, Quasiparticle relation between bulk gap nodes and surface bound states: Application to iron-based superconductors, *Phys. Rev. B* **86**, 094512 (2012).

28 S. A. Yang et al, Dirac and Weyl Superconductors, *Phys. Rev. Lett.* **113**, 046401 (2014).

29 V. Kozii, J. W. F. Venderbos, & L. Fu. Three-dimensional majorana fermions in chiral superconductors. *Sci. Adv.* **2**, e1601835 (2016).

30 F. Lambert et al., Surface State Tunneling Signatures in the Two-Component Superconductor $UPt_3$, *Phys. Rev. Lett*. **118**, 087004 (2017).

31 S. Tamura et al., Theory of surface Andreev bound states and tunneling spectroscopy in three-dimensional chiral superconductors, *Phys. Rev. B* **95**, 104511 (2017).

32 A. Crépieux, et al., Quasiparticle interference and spectral function of the $UTe_2$ superconductive surface band. arXiv:2503.17762 (2025).

33 H. Christiansen, M. Geier, B. M. Andersen, and A. Kreisel. Nodal superconducting gap structure and topological surface states of $UTe_2$, *arXiv*: 2503.11603 (2025).

34 H. Christiansen, B. M. Andersen, P. J. Hirschfeld and A. Kreisel. Quasiparticle Interference of Spin-Triplet Superconductors: Application to $UTe_2$, *arXiv*:2505.01404 (2025).

35 Q. Gu et al., Pair Wavefunction Symmetry in $UTe_2$ from Zero-Energy Surface State Visualization, *Science* **388**, 938 (2025).

36 Q.-H. Wang and D.-H. Lee, Quasiparticle scattering interference in high-temperature superconductors, *Phys. Rev. B* **67**, 020511 (2003).

37 L. Capriotti, D. J. Scalapino, and R. D. Sedgewick, Wave-vector power spectrum of the local tunneling density of states: Ripples in a d-wave sea, *Phys. Rev. B* **68**, 014508 (2003).

38 J. E. Hoffman et al., Imaging Quasiparticle Interference in $Bi_2Sr_2CaCu_2O_{8+\delta}$, *Science* **297**, 5584 1148-1151 (2002).

39 T. Hanaguri et al, Quasiparticle interference and superconducting gap in $Ca_{2-x}Na_xCuO_2Cl_2$, *Nat. Phys.* **3**, 865-871 (2007).

40 M. P. Allan et al., Anisotropic Energy Gaps of Iron-Based Superconductivity from Intraband Quasiparticle Interference in LiFeAs, *Science* **336**, 563-567(2012).

41 M. P. Allan et al., Imaging Cooper pairing of heavy fermions in $CeCoIn_5$, *Nat. Phys.* **9**, 468–473 (2013).

42 P. O. Sprau et al., Discovery of orbital-selective Cooper pairing in FeSe, *Science* **357**, 6346 (2017).

43 R. Sharma et al., Momentum-resolved superconducting energy gaps of $Sr_2RuO_4$ from quasiparticle interference imaging, *Proc. Natl. Acad. Sci. U.S.A* **117**, 5222-5227 (2020).

44 V. Kaladzhyan and C. Bena. Obtaining Majorana and other boundary modes from the metamorphosis of impurity-induced states: Exact solutions via the T-matrix. *Phys. Rev. B* **100**, 081106 (2019).

45 S. Pinon, V. Kaladzhyan, and C. Bena. Surface Green's functions and boundary modes using impurities: Weyl semimetals and topological insulators. *Phys. Rev. B* **101**, 115405 (2020).

46 M. Alvarado et al. Boundary Green's function approach for spinful single-channel and multichannel Majorana nanowires. *Phys. Rev. B* **101**, 094511 (2020).

47 H. Yoon et al., Probing *p*-wave superconductivity in $UTe_2$ via point-contact junctions. *npj Quantum Materials* **9**, 91 (2024).

Fig.1

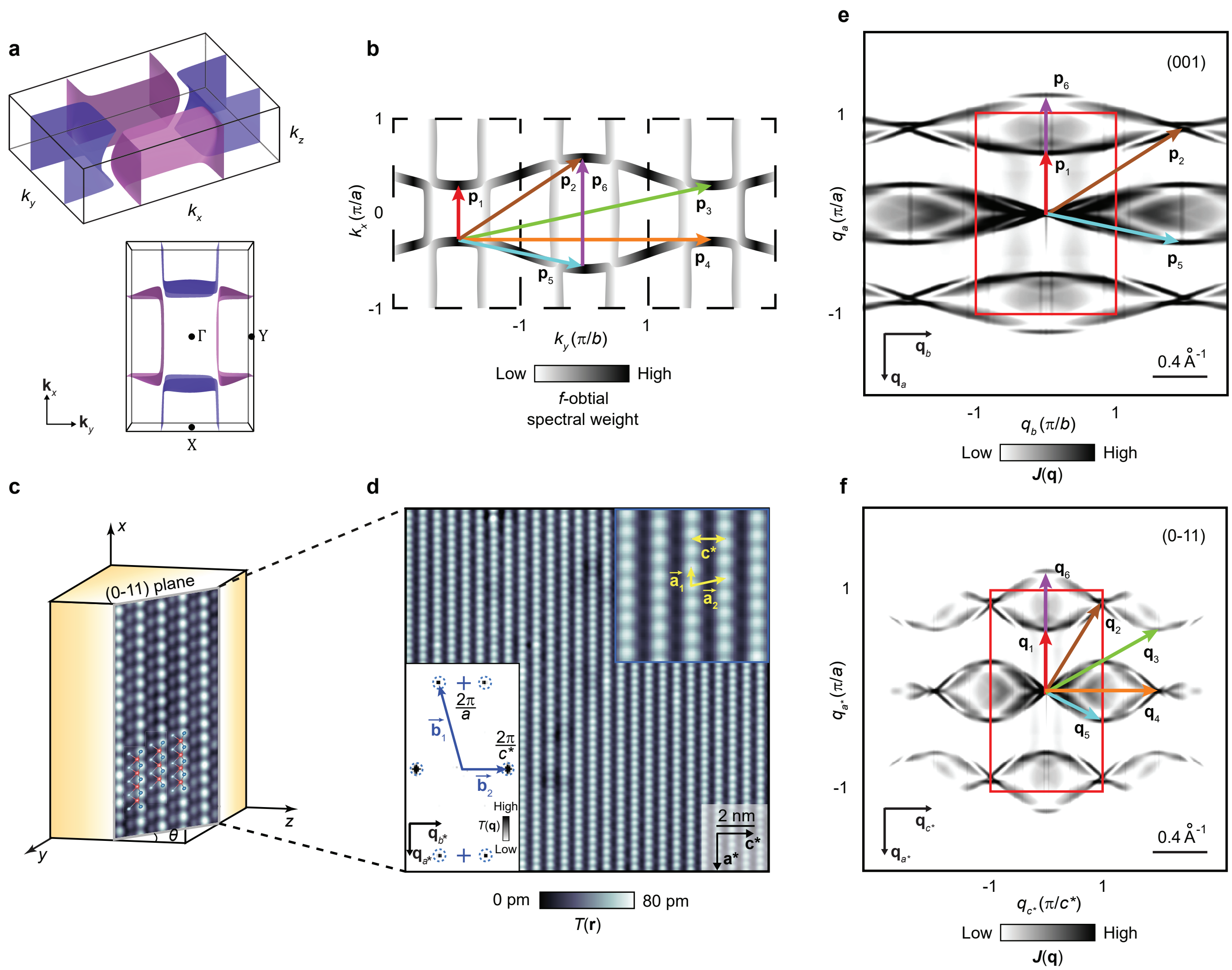

a
$k_z$
$k_y$
$k_x$
$\mathbf{k}_x$
$\mathbf{k}_y$
Γ
Y
X
b
$k_x\,(\pi/a)$
1
0
-1
-1
1
$k_y\,(\pi/b)$
$\mathbf{p}_1$
$\mathbf{p}_2$
$\mathbf{p}_3$
$\mathbf{p}_4$
$\mathbf{p}_5$
$\mathbf{p}_6$
Low High
*f*-obtial
spectral weight
e
(001)
$q_a\,(\pi/a)$
1
-1
-1
1
$q_b\,(\pi/b)$
$\mathbf{q}_b$
$\mathbf{q}_a$
0.4 Å$^{-1}$
Low High
$\mathbf{J}(\mathbf{q})$
c
x
(0-11) plane
z
y
θ
d
$\vec{\mathbf{b}}_1$
$\vec{\mathbf{b}}_2$
$\frac{2\pi}{a}$
$\frac{2\pi}{c^*}$
High
$T(\mathbf{q})$
Low
$\mathbf{q}_{b^*}$
$\mathbf{q}_{a^*}$
c*
$\vec{\mathbf{a}}_1$
$\vec{\mathbf{a}}_2$
2 nm
c*
a*
0 pm
80 pm
$T(\mathbf{r})$
f
(0-11)
$q_{a^*}\,(\pi/a)$
1
-1
-1
1
$q_{c^*}\,(\pi/c^*)$
$\mathbf{q}_1$
$\mathbf{q}_2$
$\mathbf{q}_3$
$\mathbf{q}_4$
$\mathbf{q}_5$
$\mathbf{q}_6$
$\mathbf{q}_{c^*}$
$\mathbf{q}_{a^*}$
0.4 Å$^{-1}$
Low High
$\mathbf{J}(\mathbf{q})$

Fig.2

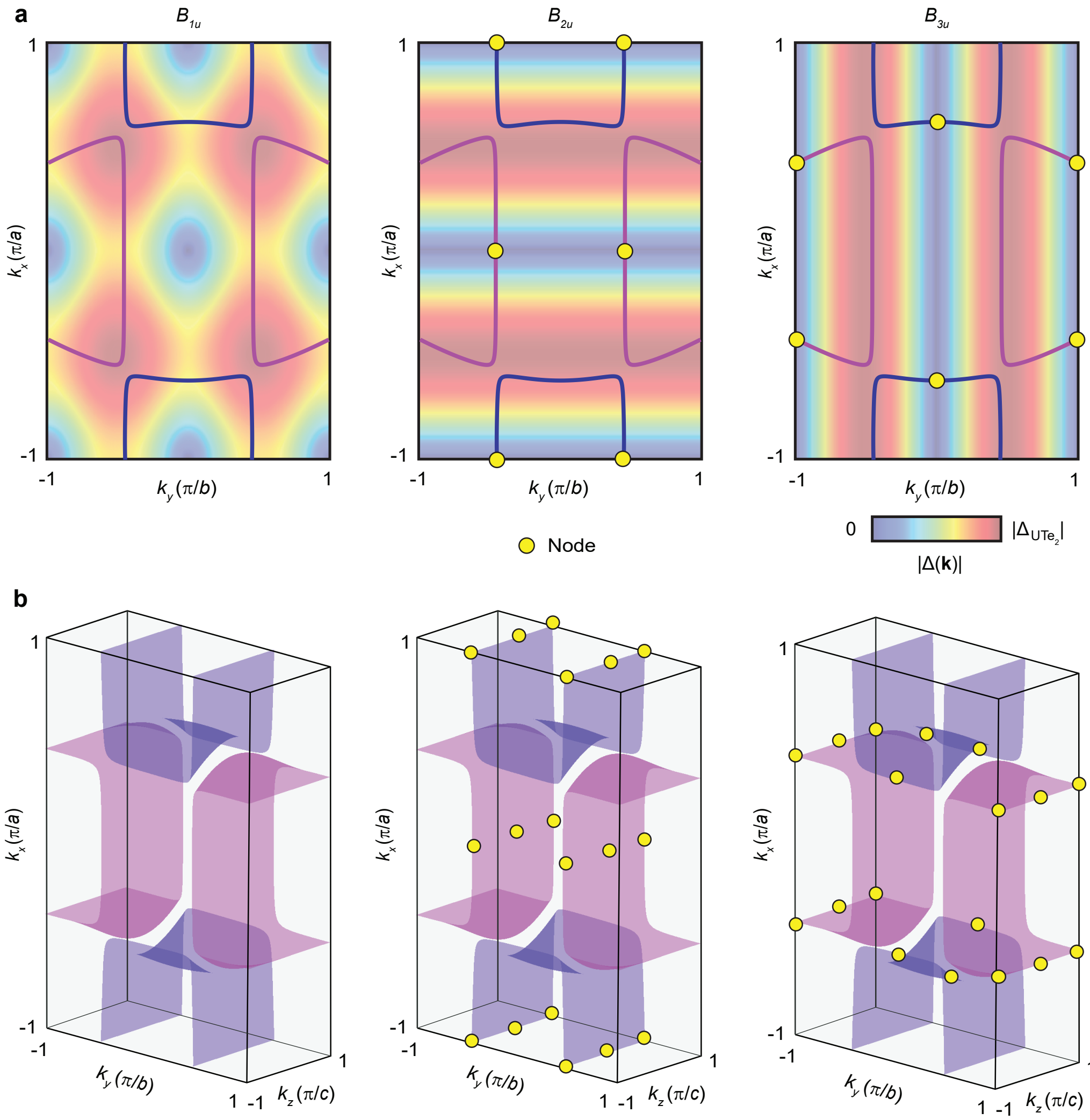

Fig.3

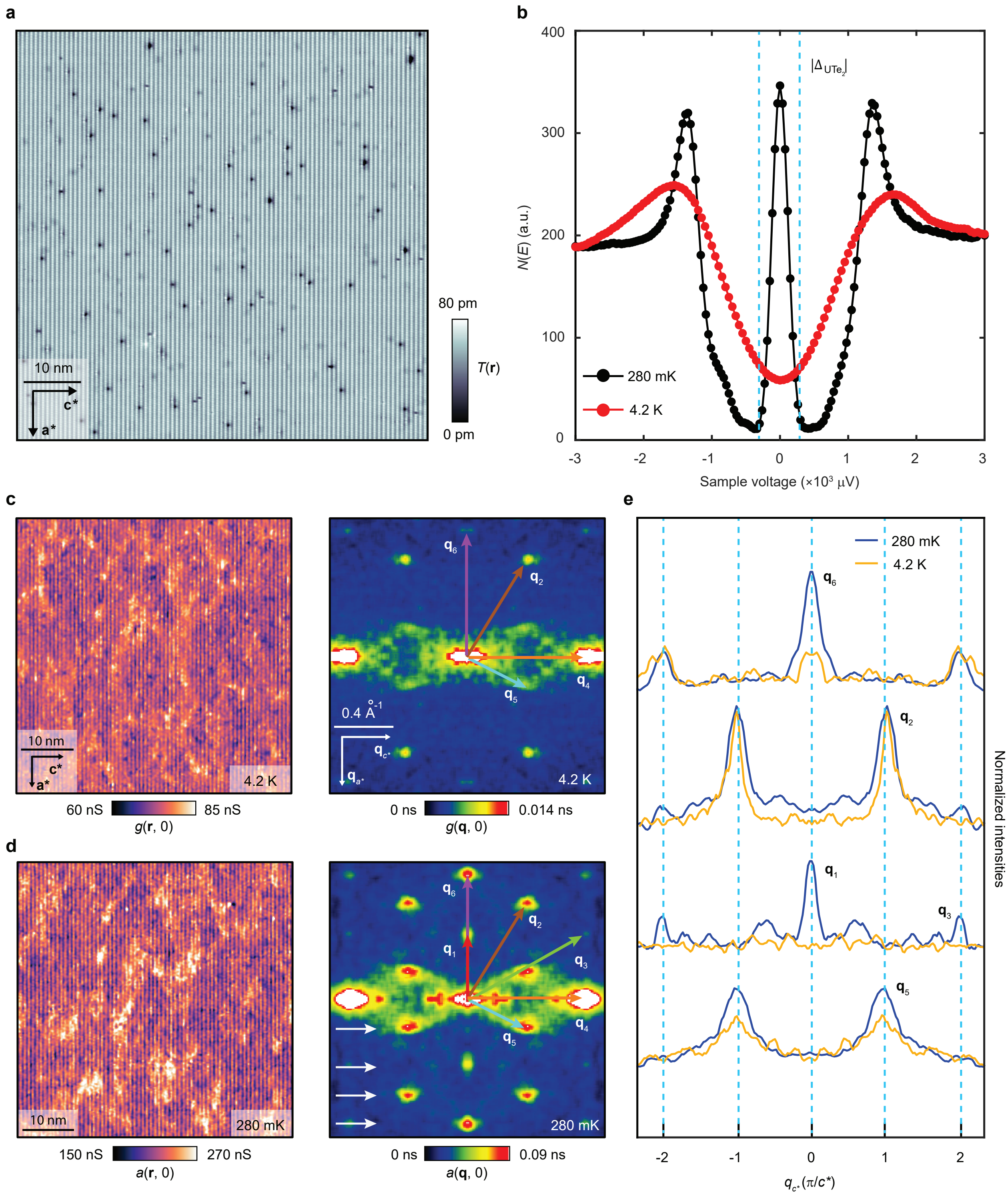

a
10 nm
c*
a*
80 pm
T(r)
0 pm
b
400
300
200
100
0
N(E) (a.u.)
|Δ UTe2|
280 mK
4.2 K
-3
-2
-1
0
1
2
3
Sample voltage (×10³ µV)
c
10 nm
c*
a*
4.2 K
60 nS
85 nS
g(r, 0)
q6
q2
q4
q5
0.4 Å-1
qc*
qa*
4.2 K
0 ns
0.014 ns
g(q, 0)
d
10 nm
280 mK
150 nS
270 nS
a(r, 0)
q6
q2
q1
q3
q4
q5
280 mK
0 ns
0.09 ns
a(q, 0)
e
280 mK
4.2 K
q6
q2
q1
q3
q5
Normalized intensities
-2
-1
0
1
2
qc* (π/c*)

Fig.4

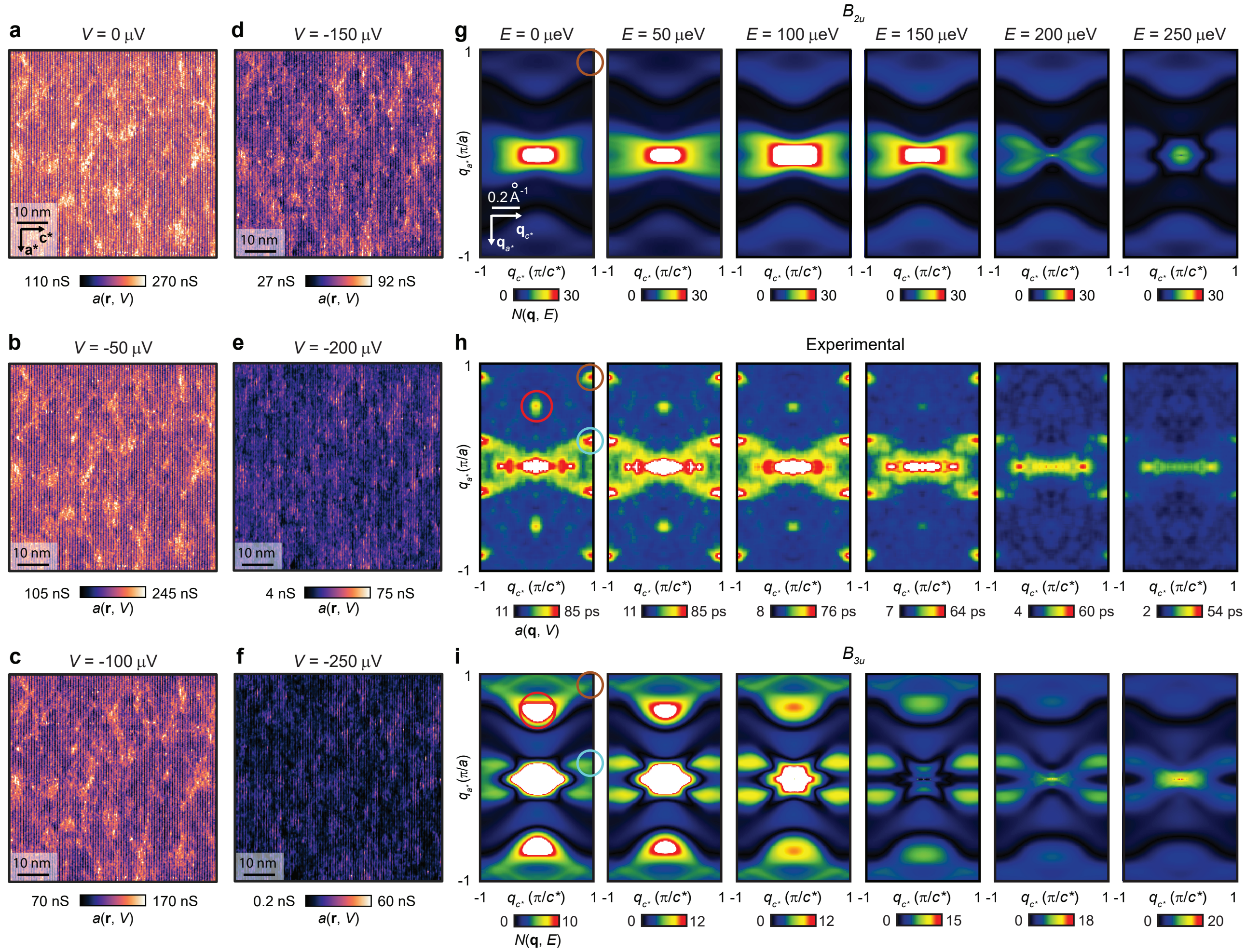

a
V = 0 μV
10 nm
c*
a*
110 nS
270 nS
a(r, V)
b
V = -50 μV
10 nm
105 nS
245 nS
a(r, V)
c
V = -100 μV
10 nm
70 nS
170 nS
a(r, V)
d
V = -150 μV
10 nm
27 nS
92 nS
a(r, V)
e
V = -200 μV
10 nm
4 nS
75 nS
a(r, V)
f
V = -250 μV
10 nm
0.2 nS
60 nS
a(r, V)
B2u
g
E = 0 μeV
E = 50 μeV
E = 100 μeV
E = 150 μeV
E = 200 μeV
E = 250 μeV
qa* (π/a)
0.2 Å-1
qc*
qa*
qc* (π/c*)
0
30
N(q, E)
h
Experimental
11
85 ps
11
85 ps
8
76 ps
7
64 ps
4
60 ps
2
54 ps
a(q, V)
i
B3u
0
10
0
12
0
12
0
15
0
18
0
20
N(q, E)

Methods for

# Odd-Parity Quasiparticle Interference in the Superconductive Surface State of $UTe_2$

## Materials and Methods

### $UTe_2$ normal state electronic structure model

In this section we first consider a 4-band tight-binding model reproducing the quasi-rectangular Fermi surface (FS) of $UTe_2$ and its undulations along $k_z$ axis, as outlined in Ref.[48]. The characteristic features are assumed to arise from the hybridization between two quasi-one-dimensional chains: one originating from the Te(2) 5$p$ orbitals and the other from the U 6$d$ orbitals. The lattice constants are taken to be $a = 0.41$ nm, $b = 0.61$ nm, $c = 1.39$ nm.

The coupling between the two Uranium orbitals is modeled by the following Hamiltonian:

$$H_{U-U} = \begin{bmatrix} \mu_U - 2t_U \cos k_x a - 2t_{ch,U} \cos k_y b & -\Delta_U - 2t'_U \cos k_x a - 2t'_{ch,U} \cos k_y b - 4t_{z,U} e^{-ik_z c/2} \cos k_x \frac{a}{2} \cos k_y \frac{b}{2} \\ -\Delta_U - 2t'_U \cos k_x a - 2t'_{ch,U} \cos k_y b - 4t_{z,U} e^{ik_z c/2} \cos k_x \frac{a}{2} \cos k_y \frac{b}{2} & \mu_U - 2t_U \cos k_x a - 2t_{ch,U} \cos k_y b \end{bmatrix} \quad (1)$$

Here the tight binding parameters are the chemical potential $\mu_U$, the intra-dimer overlap $\Delta_U$ of the uranium dimers (where two uranium atoms are coupled along the $c$-axis and the dimers run along the $a$-axis), the hopping $2t_U$ along the uranium chain in the a direction, the hopping $t'_U$ to other uranium in the dimer along the chain direction, the hoppings $t_{ch,U}$ and $t'_{ch,U}$ between chains in the $a-b$ plane, and the hopping $t_{z,U}$ between chains along the $c$-axis.

Similarly, the coupling between the two tellurium orbitals is given by:

$$H_{T_e-T_e} = \begin{bmatrix} \mu_{Te} - 2t_{ch,Te} \cos k_x a & -\Delta_{Te} - t_{Te} e^{-ik_y b} - 2t_{z,Te} \cos k_z \frac{c}{2} \cos k_x \frac{a}{2} \cos k_y \frac{b}{2} \\ -\Delta_{Te} - t_{Te} e^{ik_y b} - 2t_{z,Te} \cos k_z \frac{c}{2} \cos k_x \frac{a}{2} \cos k_y \frac{b}{2} & \mu_{Te} - 2t_{ch,Te} \cos k_x a \end{bmatrix} \quad (2)$$

where the Te tight-binding parameters are the chemical potential $\mu_{Te}$, the intra-unit-cell overlap $\Delta_{Te}$ between the two Te(2) atoms along the chain direction, the hopping $t_{Te}$ along the Te(2) chain in the $b$ direction, the hopping $t_{\mathrm{ch,Te}}$ between chains in the $a$-direction, and the hopping $t_{z,Te}$ between chains along the $c$ axis.

The hybridization between the uranium and tellurium orbitals is given by:

$$H_{U-Te} = \begin{pmatrix} \delta & 0 \\ 0 & \delta \end{pmatrix} \tag{3}$$

The normal state tight-binding Hamiltonian of $UTe_2$ can thus be written as:

$$H_{UTe_2} = \begin{pmatrix} H_{U-U} & H_{U-Te} \\ H_{U-Te}^{+} & H_{Te-Te} \end{pmatrix} \tag{4}$$

We consider the following values for the tight-binding parameters (all parameter values are expressed in units of eV): $\mu_U = -0.355$, $\Delta_U = 0.38$, $t_U = 0.17$, $t'_U = 0.08$, $t_{ch,U} = 0.015$, $t'_{ch,U} = 0.01$, $t_{z,U} = -0.0375$, $\mu_{Te} = -2.25$, $\Delta_{Te} = -1.4$, $t_{Te} = -1.5$, 0, $t_{ch,Te} = 0$, $t_{z,Te} = -0.05$, $\delta = 0.13$. These parameters are chosen to be consistent with both quantum oscillation measurements and our QPI data. All the hopping terms considered here are between the two nearest neighbours such that all scattering will be constrained to nearest neighbour sites at the surface. Any impurity potential is taken to be fully diagonal in the orbital basis with equal intensity on U-orbitals and Te-orbitals. These parameters are used in all simulations presented herein.

### $UTe_2$ superconductive energy gap nodes and their (0-11) projections

Nodal locations presented in the main text are derived from the general expression for the electronic dispersion of a spin-triplet superconductor[6]

$$E_{\mathbf{k}}^{\pm} = \sqrt{\varepsilon^2(\mathbf{k}) + |\mathbf{d}(\mathbf{k})|^2 \pm |\mathbf{d}(\mathbf{k}) \times \mathbf{d}^*(\mathbf{k})|} \tag{5}$$

where $\varepsilon(\mathbf{k})$ is the normal state dispersion measured from the chemical potential and $\mathbf{d}(\mathbf{k})$ is the **d**-vector order parameter. The gap functions we have considered are those associated with the odd-parity irreducible representations (IRs) of the point group $D_{2h}$, namely,

| IR | d-vector |
|---|---|
| $A_u$ | $[C_1\sin(k_x a)\,, C_2\sin(k_y b)\,, C_3\sin(k_z c)]$ |
| $B_{1u}$ | $[C_1\sin(k_y b)\,, C_2\sin(k_x a)\,, C_0\sin(k_x a)\sin(k_y b)\sin(k_z c)]$ |
| $B_{2u}$ | $[C_1\sin(k_z c)\,, C_0\sin(k_x a)\sin(k_y b)\sin(k_z c)\,, C_3\sin(k_x a)]$ |
| $B_{3u}$ | $[C_0\sin(k_x a)\sin(k_y b)\sin(k_z c)\,, C_2\sin(k_z c)\,, C_3\sin(k_y b)]$ |

**Table 1**. Odd-parity irreducible representations of the crystal point symmetry group $D_{2h}$ and the corresponding **d**-vectors representations for the simple orthorhombic lattice model used throughout this paper.

In all cases $\mathbf{d}(\mathbf{k}) = \mathbf{d}^*(\mathbf{k})$, the gap function is unitary and the nodal locations are defined by FS intersections with the high-symmetry lines of the Brillouin zone Within this model, the nodal points are indicated by yellow dots in Extended Data Figs. 1a,b,c for $B_{1u}$, $B_{2u}$, and $B_{3u}$ respectively. For $B_{1u}$ symmetry, the FS is fully gapped. While sharing the same number of independent nodes, the locations of the nodes are extremely different in the 3D Brillion zone for the $B_{2u}$ and $B_{3u}$ order parameters (Extended Data Figs. 1b-c).

Next, we project the normal state FS onto the (0-11) plane oriented at an angle of 24° between the normal to the (0-11) plane and crystal $b$-axis (Extended Data Fig. 1d). The result is a (0-11) surface Brillouin zone (SBZ). The basis vectors on this (0-11) plane are $\mathbf{e}_a = (1,0,0)$ and $\mathbf{e}_{c^*} = (0, \sin\theta, cos\theta)$ where $\theta = 24°$. When an arbitrary vector of $(a, b, c)$ is projected to the (0-11) plane, the projected vector is $\left((a, b, c)\cdot\mathbf{e}_a,\ (a, b, c)\cdot\mathbf{e}_{c^*}\right) = (a, 0.4b + 0.91c)$. This occurs because any momentum $\mathbf{k}$ of the bulk BZ can be decomposed into momentum components parallel to the plane $\mathbf{k}_\parallel$ and components perpendicular to the plane $\mathbf{k}_\perp$, of the surface. Then only $\mathbf{k}_\parallel$ will contribute to the surface quasiparticle states as $\mathbf{k}_\perp$is no longer a conserved quantity i.e. the (001) quasiparticle states that are transformed into $\mathbf{k}_\perp$ states in (0-11) plane no longer contribute. This is why the scale of **q**-space and the size of the SBZ are both reduced when viewed at the (0-11) termination surface of $UTe_2$.

Finally, we project the bulk nodes onto the (0-11) plane and obtain a **k**-space projected-nodal structure for order parameters $B_{1u}$, $B_{2u}$, and $B_{3u}$ respectively (Extended Data Figs. 1e-g). By definition $A_u$ and $B_{1u}$ have no bulk or projected energy-gap nodes and so we consider them no further. However, at the (0-11) SBZ of $UTe_2$ the projected nodal locations of the bulk $B_{2u}$ order parameter are fundamentally different from those of the bulk $B_{3u}$ order parameter as shown in Extended Data Figs. 1f and 1g respectively.

**Quasiparticle scattering interference in the QSB at the (0-11) surface of $UTe_2$**

We choose to work in the following basis, where $U_{1/2}$ and $Te_{1/2}$ denote respectively the two uranium and tellurium orbitals:

$$\psi^+(\mathbf{k}) = (c^+_{U_1,\mathbf{k},\sigma}\,, c^+_{U_2,\mathbf{k},\sigma}\,, c^+_{Te_1,\mathbf{k},\sigma}\,, c^+_{Te_2,\mathbf{k},\sigma}\,, c_{U_1,-\mathbf{k},\bar{\sigma}}\,, c_{U_2,-\mathbf{k},\bar{\sigma}}\,, c_{Te_1,-\mathbf{k},\bar{\sigma}}\,, c_{Te_2,-\mathbf{k},\bar{\sigma}}) \quad (6)$$

$$c^+_{\alpha,\mathbf{k},\sigma} = (c^+_{\alpha,\mathbf{k},\uparrow}\,, c^+_{\alpha,\mathbf{k},\downarrow}) \quad (7)$$

$$c_{\alpha,\mathbf{k},\bar{\sigma}} = (c_{\alpha,\mathbf{k},\downarrow}\,, c_{\alpha,\mathbf{k},\downarrow}) \quad (8)$$

In this basis the BdG Hamiltonian of a *p*-wave spin triplet superconductor can be written as:

$$H_{BdG}(\mathbf{k}) = \psi^+(\mathbf{k}) \begin{pmatrix} H_{\mathrm{UTe_2}}(\mathbf{k})\otimes I_2 & \Delta(\mathbf{k})\otimes I_4 \\ \Delta^+(\mathbf{k})\otimes I_4 & -H^*_{\mathrm{UTe_2}}(-\mathbf{k})\otimes I_2 \end{pmatrix} \psi(\mathbf{k}) \quad (9)$$

where the order parameter for the putative *p*-wave superconductor is $\Delta(\mathbf{k}) = \Delta_0 i(\mathbf{d}\cdot\boldsymbol{\sigma})\sigma_2$, $I_n$ is an $n \times n$ identity matrix. In our analysis we focus on the non-chiral order parameters: $A_u$, $B_{1u}$, $B_{2u}$, and $B_{3u}$. The **d**-vectors used in calculations for each IR are provided in Methods Table 1.

In our simulations we hypothesize the following values: $C_0 = 0$, $C_1 = 300$ μeV, $C_2 = 300$ μeV, and $C_3 = 300$ μeV. In this conventional model $C_1$, $C_2$ and $C_3$ are hypothesized to be the same as the $UTe_2$ gap amplitude measured in experiment. Although the relative intensity of these coefficients is not known *a priori* we have checked that, while keeping the maximum gap constant, these coefficient values produce the same QPI features with only slight changes in wavevector length. Within this model, the unperturbed retarded bulk three-dimensional Green's function is given as:

$$G_0(\boldsymbol{k}, \omega) = [(\omega + i\eta)I - H_{BdG}(\boldsymbol{k})]^{-1} \quad (10)$$

with the corresponding unperturbed spectral function written as:

$$A_0(\boldsymbol{k}, \omega) = -1/\pi \, \mathrm{Im}\, G_0(\boldsymbol{k}, \omega) \quad (11)$$

where $\eta$ is the energy broadening factor in the theory simulation.

Although obtaining the bulk Green's function is straightforward, calculating the surface Green's functions and spectral functions $A_s(\mathbf{k}, \omega)$ is significantly more difficult. The complexity arises because the surface Green's functions characterize a semi-infinite system with broken translational symmetry, thus they cannot be calculated directly. Traditionally they are obtained using heavy numerical recursive Green's function techniques as in Ref. [49]. Here we use a novel and simpler analytical technique, described in references [44-46], in which the surface is modeled using a planar impurity. When the magnitude of the impurity potential goes to infinity, the impurity splits the system into two semi-infinite spaces. Then only wavevectors in the (0-11) plane remain good quantum numbers. The effect of this impurity can be exactly calculated using the T-matrix formalism, which gives one access to the surface Green's function of the semi-infinite system.

We model the effect of the surface using a planar-impurity-potential as in Extended Data Fig. 2, which is oriented parallel to the (0-11) crystal plane. In the presence of this impurity, the bulk Green's function is modified to

$$G(\mathbf{k}_1, \mathbf{k}_2, \omega) = G_0(\mathbf{k}_1, \omega)\delta_{\mathbf{k}_1,\mathbf{k}_2} + G_0(\mathbf{k}_1, \omega)T(\mathbf{k}_1, \mathbf{k}_2, \omega)G_0(\mathbf{k}_2, \omega) \quad (12)$$

where the T matrix considers all-order impurity scattering processes. For a plane impurity localized at $x = 0$ and perpendicular to the $x$ axis, the T matrix can be computed as

$$T(k_{1y}, k_{1z}, k_{2y}, k_{2z}, \omega) = \delta_{k_{1y},k_{2y}}\delta_{k_{1z},k_{2z}}[1 - \hat{V}\int \frac{dk_x}{L_x} G_0(k_x, k_{1y}, k_{1z}, \omega)]^{-1}\hat{V} \quad (13)$$

with $L_x$ a normalization factor. Since the impurity potential is a delta function in $x$, the T-matrix is independent on $k_x$, and depends only on $k_y$ and $k_z$.

We calculate the exact Green's function one lattice spacing away from the planar-impurity-potential, which converges precisely to the surface Green's function as the impurity potential approaches infinity. This surface Green's function can be obtained by performing a partial Fourier transform of the exact Green's function expressed in Eq. (12)

$$G_s(k_y, k_z) = \int \frac{dk_{1x}}{L_x} \int \frac{dk_{2x}}{L_x} G(k_{1x}, k_y, k_z, k_{2x}, k_y, k_z, \omega) e^{ik_{1x}x} e^{-ik_{2x}x'} \qquad (14)$$

where $x = x' = \pm 1$.

Extended Data Figs. 3a-c are generated using the above (0-11) planar-impurity-potential formalism for the 4-band model with $B_{1u}$, $B_{2u}$, and $B_{3u}$ gap structures. In Extended Data Fig. 3 we present the surface spectral function $A_s(\mathbf{k}, E)$ for these order parameters in the (0-11) SBZ. In particular, the surface spectral function $A_s(\mathbf{k}, E)$ for $B_{3u}$ in the (0-11) SBZ is shown in Extended Data Fig. 3c. A hypothesized sextet of scattering wavevectors $\mathbf{q}_i$, $i$=1-6 connecting regions of maximum intensity in $A_s(\mathbf{k}, 0)$ is overlaid. All plots show data for six energy levels, with the highest near the gap edge of $|\Delta_{UTe_2}| = 300$ μeV.

We next describe how QPI scattering is possible given the putative protection of superconductive topological surface band quasiparticles against scattering in a topological superconductor. Formally, we can derive the spin-resolved quasiparticle surface spectral function as shown for a $B_{2u}$ and $B_{3u}$ QSB in Extended Data Fig. 4. The resulting surface spectral function can be clearly segregated into two spin-polarized bands in $UTe_2$, one for each spin eigenstate. While spin-flip and thus inter-spin-band scattering is proscribed, non-spin-flip or intra-spin-band scattering is allowed thus allowing QPI of these quasiparticles.

Extended Data Figs. 5a,b depict the projection of the bulk spectral function of order parameters $B_{2u}$ and $B_{3u}$ on the (0-11) surface. It should be noted that the resulting features correspond to regions identifiable from the 3D bulk FS as the projection of the bulk nodes

onto the (0-11) surface and these features are highlighted by yellow circles. Extended Data Figs. 5c,d depict the surface spectral function $A_s(\mathbf{k}, 0)$ computed using the planar-impurity method[44]. It accounts for some bulk contributions but is dominated by new features which connect the projection of the bulk nodes to the SBZ, these new features correspond to the QSB of order parameters $B_{2u}$ and $B_{3u}$.

In Extended Data Figs. 5e,f we consider (0-11) surface QPI featuring order parameters of $B_{2u}$ and $B_{3u}$ symmetry using the joint density of states (JDOS) $J(\boldsymbol{q}, 0)$. The JDOS approximation $J(\mathbf{q}, 0)$ is a well-established technique to map out the geometries of the momentum-space band structures[36]. The JDOS approximation is based on the observation that if the surface spectral function $A_s$ at $\mathbf{k}$ and $\mathbf{k} + \mathbf{q}$ are both simultaneously large then $J(\mathbf{q}, E)$ will be large as $\mathbf{q}$ connects regions of large joint density of states. This technique has been used to successfully interpret the experimental QPI data for high temperature superconductors[50,51], topological insulators[52,53], and Weyl semimetals[54,55].

Although $J(\mathbf{q}, E)$ captures the dominant $\mathbf{k}$-space quasiparticle scattering associated with the order parameter symmetries it does not consider spin forbidden scattering processes and the underlying contributions from the bulk band structure as accurately as the $N(\mathbf{q}, E)$ simulations presented in the main text. However, both $J(\mathbf{q}, E)$ and $N(\mathbf{q}, E)$ calculations reveal distinct scattering features.

We show the theoretical $N(E)$ calculations for the $UTe_2$ (0-11) surface with $B_{2u}$ and $B_{3u}$ gap symmetries in Extended Data Figs. 5g-h. Both gap symmetries show the indistinguishable bulk $N(E)$ of a nodal $p$-wave superconductor (black curve). The $N(E)$ at the surface (red curve) differs significantly between the two order parameter symmetries in this model. For a $B_{3u}$ order parameter the surface $N(E)$ has a clear zero-energy peak however the surface $N(E)$ due to a $B_{2u}$ order parameter has only reduced gap depth compared to bulk. In experiment, we find intense zero-energy conductance, which appears most consistent with the (0-11) surface $N(E)$ in the presence of $B_{3u}$ gap symmetry.

To further improve the comparison between the QPI simulations and the experimental QPI data, we consider of the **q**-space sensitivity of our scan tip in the QPI simulations. The QPI simulations $N(\mathbf{q}, E)$ for the $B_{2u}$ and $B_{3u}$ order parameters are shown in Extended Data Figs. 6a-b, which show very strong intensities near the high-**q** region. In experimental data, however, the intensity near the high-**q** regions which represent shortest distances in **r**-space, decays rapidly due to the finite radius of the scan tip. We estimate the actual **q**-space intensity decay radius from a gaussian fit to the power spectral density of the relevant $T(\mathbf{q})$ image. Subsequently we apply a 2D Gaussian function of the following form to the QPI simulations $N(\mathbf{q}, E)$, reflecting the effects of the finite circular radius or 'aperture' of the scan tip

$$f(q_x, q_y) = Aexp\left(-\left(\frac{(\boldsymbol{q}_x - \boldsymbol{q}_{x_0})^2}{2\sigma_x^2} + \frac{(\boldsymbol{q}_y - \boldsymbol{q}_{y_0})^2}{2\sigma_y^2}\right)\right) \tag{15}$$

where the amplitude $A = 1.75 \times 10^{-5}$, the center coordinates $(q_{x_0}, q_{y_0}) = (0, 0)$ and the standard deviation $\sigma_x = \sigma_y = 3.68\pi/c^*$. Upon applying this 2D 'aperture' filter in Extended Data Figs. 6a-b, we derive the $N(\mathbf{q}, E)$ in main text Figs. 4g and 4i.

To evaluate the effect of impurity strength on the QPI calculations, we have performed superconductive topological surface band QPI simulations using local impurity potentials of $V$ = 0.07 eV, 0.2 eV, 0.5 eV and 1 eV potentials and find that the predictions using different scattering impurity potentials lead to highly consistent scattering wavevectors (Extended Data Fig. 6c-f). Varying the scattering potentials only changes relative amplitudes at different wavevectors; when the scattering potentials increase, the scattering wavevectors caused by the surface state become more intense. We have chosen $V$ = 0.2 eV because the QPI simulations calculated using this scattering potential are most consistent with the relative QPI intensities observed experimentally. The scientific conclusions that QPI in the superconductive topological surface state of $UTe_2$ is consistent with $B_{3u}$ bulk pairing symmetry, remain unchanged when using scattering potentials ranging from 0.07 eV to 1 eV as presented above.

**Surface Andreev bound state in unconventional superconductors**

The surface Andreev bound state (SABS) and concomitant zero-bias conductance peaks due to π phase shifts have been extensively studied for decades, particularly in high temperature superconductors[56-60]. In *d*-wave superconductors such as the cuprates, the π phase shift of the pair potential occurs universally when the angle between the crystal axis of the superconductors and the lobe direction of *d*-wave pair potential is nonzero. This phase shift leads to the formation of SABS due to Andreev reflection. These SABS manifest as zero-bias conductance peaks (ZBCP) in tunneling spectroscopy, a hallmark feature widely observed and investigated in the cuprate high temperature superconductors.

Although never observed experimentally, Andreev zero bias conductance should emerge in three-dimensional *p*-wave intrinsic topological superconductors, where they are often described as superconducting topological surface states. These SABS have a somewhat distinct physical origin with those in *d*-wave systems because, in an odd-parity superconductors, there is a universal π phase shift of the superconducting order parameter at all surfaces independent from the angle between the crystal axis and the direction of the phase of the superconducting order parameter.

**Alternative Gap Function and Impurity Potential**

Owing to the body-centered orthorhombic crystal symmetry of $UTe_2$, basis functions other than those presented in the main text and above are allowed. To consider alternative basis functions, we add additional, symmetry-allowed, terms to the **d**-vectors as described in Ref. 8. For the nodal, single-component order parameters we then use the below d-vectors with $C_0 = 0$, $C_1 = C_2 = C_3 = 0.225$ meV, and $C_4 = C_5 = C_6 = 0.15$ meV.

| | |
|---|---|
| $B_{2u}$ | $\begin{pmatrix} C_1\sin(k_zc)+C_4\sin\frac{k_zc}{2}\cos\frac{k_xa}{2}\cos\frac{k_yb}{2} \\ C_0\sin(k_xa)\sin(k_yb)\sin(k_zc) \\ C_3\sin(k_xa)+C_6\sin\frac{k_xa}{2}\cos\frac{k_yb}{2}\cos\frac{k_zc}{2} \end{pmatrix}$ |
| $B_{3u}$ | $\begin{pmatrix} C_0\sin(k_xa)\sin(k_yb)\sin(k_zc) \\ C_2\sin(k_zc)+C_5\sin\frac{k_zc}{2}\cos\frac{k_xa}{2}\cos\frac{k_yb}{2} \\ C_3\sin(k_yb)+C_6\sin\frac{k_yb}{2}\cos\frac{k_xa}{2}\cos\frac{k_zc}{2} \end{pmatrix}$ |

**Table 2**. The **d**-vectors representations for the body centered orthorhombic lattice model.

To establish that conclusions derived in the main text would be unchanged if these alternative **d**-vectors were used, we calculate the bulk projected spectral function $A_0(\mathbf{k}, E)$, surface spectral function $A_s(\mathbf{k}, E)$, and $J(\mathbf{q}, E)$ using these alternative triplet $\boldsymbol{d}$-vectors. These data are presented in Extended Data Fig. 7 for $E = 0$. The nodal pattern highlighted with yellow dashed circles in Extended Data Figs. 7a,b can be directly compared to Extended Data Figs. 5a,b. The alternative **d**-vectors have a very similar nodal pattern when projected to the (0-11) plane and thus the QSB occupies similar regions of the projected SBZ. This can be seen in Extended Data Figs. 7c,d in which we plot $A_s(\mathbf{k}, 0)$. From comparison with Extended Data Figs. 5c,d we see clearly that the QSB calculated using either the main text **d**-vector or these alternative **d**-vectors are nearly identical. The resulting $J(\mathbf{q}, E)$, is presented in Extended Data Figs. 7e,f for order parameter symmetries $B_{2u}$ and $B_{3u}$ respectively. Using the same quasiparticle broadening parameter as in Extended Data Figs. 5e,f, $\eta = 30$ μeV but now with these alternative **d**-vector terms, we see that the $J(\mathbf{q}, E)$ QPI patterns predicted for each order parameter have the same key features.

**Andreev conductance a(r,*V*) of quasiparticle surface band quasiparticles**

A key consideration is the role of QSB mediated Andreev conductance across the junction between *p*-wave and *s*-wave superconductors (Extended Data Fig. 8). Most simply, a single Andreev reflection transfers two electrons (holes) between the tip and the sample. Based on an S-matrix approach, the formula to compute the Andreev conductance of the $s$-wave – insulator – $p$-wave (SIP) model is

$$a(V) = \frac{8\pi^2 t_{\text{eff}}^4 e^2}{h} \sum_n \frac{\langle\phi_n|P_h|\phi_n\rangle\langle\phi_n|P_e|\phi_n\rangle}{(eV-E_n)^2+\pi^2 t_{\text{eff}}^4[\langle\phi_n|P_h|\phi_n\rangle+\langle\phi_n|P_e|\phi_n\rangle]^2} \tag{16}$$

Here $|\phi_n\rangle$ is the projection of the $n^{\text{th}}$ QSB eigenfunction onto the top $UTe_2$ surface, and $P_e$ and $P_h$ are the electron and hole projection operators acting on the $UTe_2$ surface and $V$ is the bias voltage. Thus, in principle, and as outlined in Ref. 35, superconductive scan tips can be employed as direct probes of quasiparticle surface band, with tip-sample conductance mediated by Andreev transport through the QSB.

**Distinguish between Andreev tunneling and Josephson tunneling**

Determining the physical origin of the zero bias conductance is due to Josephson or Andreev tunneling is important. However, Josephson currents are undetectable in all Nb/$UTe_2$ junctions that we have studied. This can be demonstrated by comparing the zero-bias (Andreev) conductance $a(0)$ versus junction resistance $R$ on the same plot with the maximum possible zero-bias conductance $g(0)$ which could be generated by the Josephson effect (as shown in Supplementary Fig. S6 of Ref. 35). Firstly, at high $R$, the intensity of measured $a(0)$ of Nb/$UTe_2$ junctions is orders of magnitude larger than it could possibly be due to Josephson currents (here exemplified by measured Nb/$NbSe_2$ Josephson effect zero-bias conductance[61] that itself should be at least five times larger than any which could exist in Nb/$UTe_2$). Secondly, measured $a(0)$ for Nb/$UTe_2$ first grows linearly with falling $R$ but then diminishes steeply as $R$ is reduced further. By contrast, zero-bias conductance due to Josephson currents $g(0)$ must grow rapidly and continuously as $1/R^2$ as exemplified in the Nb/$NbSe_2$ $g(0)$ data[61]. These facts (Supplementary Fig. S6 of Ref. 35) demonstrate the absolute predominance of Andreev tunneling and the non-observability of Josephson currents between Nb electrodes and the $UTe_2$ (0-11) termination surface.

**Normal-tip and superconductive-tip study of quasiparticle surface bands**

Motivated by the presence of dominant finite density of states at zero-energy as $T \to 0$ and by the consequent hypothesis that a QSB exists in this material, we searched for its signatures using a non-superconductive tip, at voltages within the superconducting energy gap, and identify unique features resulting from QSB scattering interference. The typical NIS tunneling conductance of the $UTe_2$ superconducting state measured using a non-superconductive tip is exemplified in the inset to Extended Data Fig. 9b. At the (0-11) surface of superconducting $UTe_2$ crystals almost all states inside the superconducting gap $|E| < \Delta_0$ show residual, ungapped density of states. A combination of impurity scattering and the presence of a QSB on this crystal surface are expected for a $p$-wave superconductor. Both types of these unpaired quasiparticles should contribute to conductance measurements performed within the superconducting gap using a non-superconductive scan tip. To visualize the scattering interference of QSB quasiparticles, we focus on a 40 nm square FOV (Extended Data Fig. 9a) for conventional normal-tip differential conductance $dI/dV|_{\mathrm{NIS}}(\mathbf{r}, V)$ at $T$ = 280 mK and at a junction resistance of $R$ = 5 MΩ. Although the QPI inside the superconducting gap shows some evidence of the QSB in $UTe_2$, its weak signal-to-noise ratio owing to the dominant finite density of states for $|E| \leq \Delta_0$ implies that conventional $dI/dV|_{\mathrm{NIS}}(\mathbf{q}, V)$ spectra are inadequate for precision application of detecting and quantifying the QPI of the QSB in $UTe_2$.

Thus, we turned to a new technique by using superconductive tips to increase the signal-to-noise ratio of QSB quasiparticle scattering. Recent theory for the tunnel junction formed between an $s$-wave superconductive scan-tip and a $p$-wave superconductor with a QSB within the interface[35], reveals that the high density of QSB quasiparticles allows efficient creation/annihilation of Cooper pairs in both superconductors, thus generating intense Andreev differential conductance $a(\mathbf{r}, V) \equiv dI/dV|_{\mathrm{A}}(\mathbf{r}, V)$. This is precisely what is observed when $UTe_2$ is studied by superconductive Nb-tip STM at $T$ = 280 mK, as evidenced by the large zero-energy conductance peak around $a(\mathbf{r}, 0)$ (inset to Extended Data Fig. 9d). Visualization of $a(\mathbf{r}, 0)$ and its Fourier transform $a(\mathbf{q}, 0)$ as shown in Extended Data Fig. 9d, reveals intense conductance modulations and a distinct QPI pattern. Comparing $g(\mathbf{q}, 0)$ in Extended Data Fig. 9b and $a(\mathbf{q}, 0)$ in Extended Data Fig. 9d reveals numerous common characteristics thus demonstrating that use of $a(\mathbf{q}, V)$ imaging yields equivalent QPI patterns as $g(\mathbf{q}, V)$ imaging, but with greatly enhanced signal-to-noise ratio. This is as expected since spatial variations in the intensity of $a(\mathbf{r}, V)$ are controlled by the amplitude of QSB quasiparticle wavefunctions as in Eqn. 16, so that spatial interference patterns of the QSB quasiparticles will become directly observable in $a(\mathbf{r}, V)$. Thus, visualizing spatial variations in $a(\mathbf{r}, V)$ and their Fourier transforms $a(\mathbf{q}, V)$ enables efficient, high signal-to-noise ratio, exploration of QSB quasiparticle scattering interference phenomena at the surface of $UTe_2$.

**Independent QSB visualization experiments**

To confirm that the QPI of the QSB is repeatable, we show two additional examples of the Andreev QPI $a(\mathbf{q}, 0)$ from two different FOVs in Extended Data Fig. 10. The QPI maps $a(\mathbf{q}, 0)$ are measured at zero energy where the Andreev conductance is most prominent. The two QPI $a(\mathbf{q}, 0)$ maps in Extended Data Figs. 10a,b show vividly the same sextet of scattering wavevectors $\mathbf{q}_i$, $i$ = 1-6 reported in the main text and further confirm the signatures of a $B_{3u}$-QSB in $UTe_2$. Particularly, repeated measurements of the $\mathbf{q}_1$ wavevector exclusively both within the superconducting energy gap and at $T$ = 280 mK, support the presence of a superconducting order parameter with $B_{3u}$ symmetry as this is the only order parameter which allows spin-conserved scattering at $\mathbf{q}_1$ . These two QPI maps are measured independently in two different FOVs and at two different scanning angles (Extended Data Figs. 10c, d).

**Origin of the scattering wavevector $\mathbf{q}_1$**

The interaction with uniform superconductivity of the $UTe_2$ pre-existing CDW or of the consequent PDW both occurring with the same wavevector $\mathbf{Q} = \mathbf{q}_6$, cannot induce either a CDW or a PDW at $\mathbf{Q}/2$. This is ruled out by Ginzburg-Landau theory[62]. As to the appearance of a new fundamental PDW at a $\mathbf{q}_1$, this has been ruled out previously by direct search for energy gap modulations at that wavevector[13].

The emergence of $\mathbf{q}_1$ scattering intensely in the superconducting state of $UTe_2$ occurs naturally because this wavevector arises from Bogoliubov quasiparticle scattering between symmetry-imposed superconducting nodes of the $B_{3u}$ order parameter[34]. In the normal scattering between Fermi surfaces at this wavevector may also occur, but is not predominant.

Notably the superconducting gap nodes of the $B_{3u}$ order parameter coincide with the location of the normal state Fermi surface nesting points. Consequently, the QPI wavevectors observed in the superconducting state of $UTe_2$ (Fig. 3d) coincide with the normal state Fermi surface nesting vectors. This is not necessarily the case in other superconductors such as $Sr_2RuO_4$ where the Bogoliubov QPI scattering wavevectors are entirely different from the normal-state Fermi surface nesting vectors because of the locations of the nodes in that material[43].

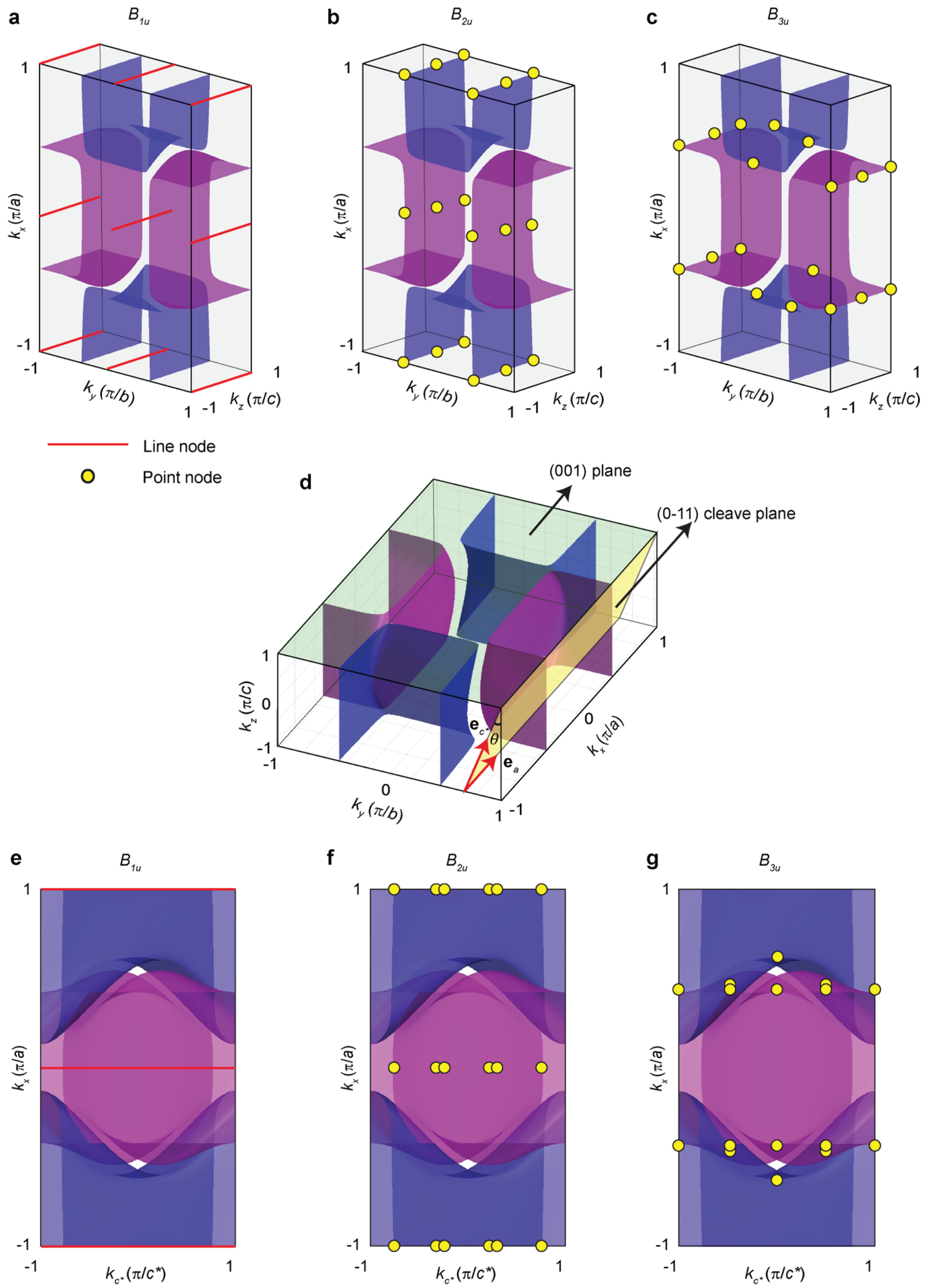

**Extended Data Fig. 1 | Projection of Fermi surfaces and gap nodes in $B_{1u}$, $B_{2u}$, and $B_{3u}$. a-c.** Bulk FS of $UTe_2$ in 3D Brillion zone, showing the nodeless $B_{1u}$ gap, eight independent $B_{2u}$ gap nodes, and eight independent $B_{3u}$ gap nodes. The red lines indicate the location where the order parameter vanishes. **d.** Projection of the (001) plane (green) onto the (0-11) plane SBZ (yellow). **e-g.** Bulk FS of $UTe_2$ projection onto the (0-11) plane SBZ, showing the $B_{1u}$ gap nodal lines, $B_{2u}$ gap nodes, and $B_{3u}$ gap nodes.

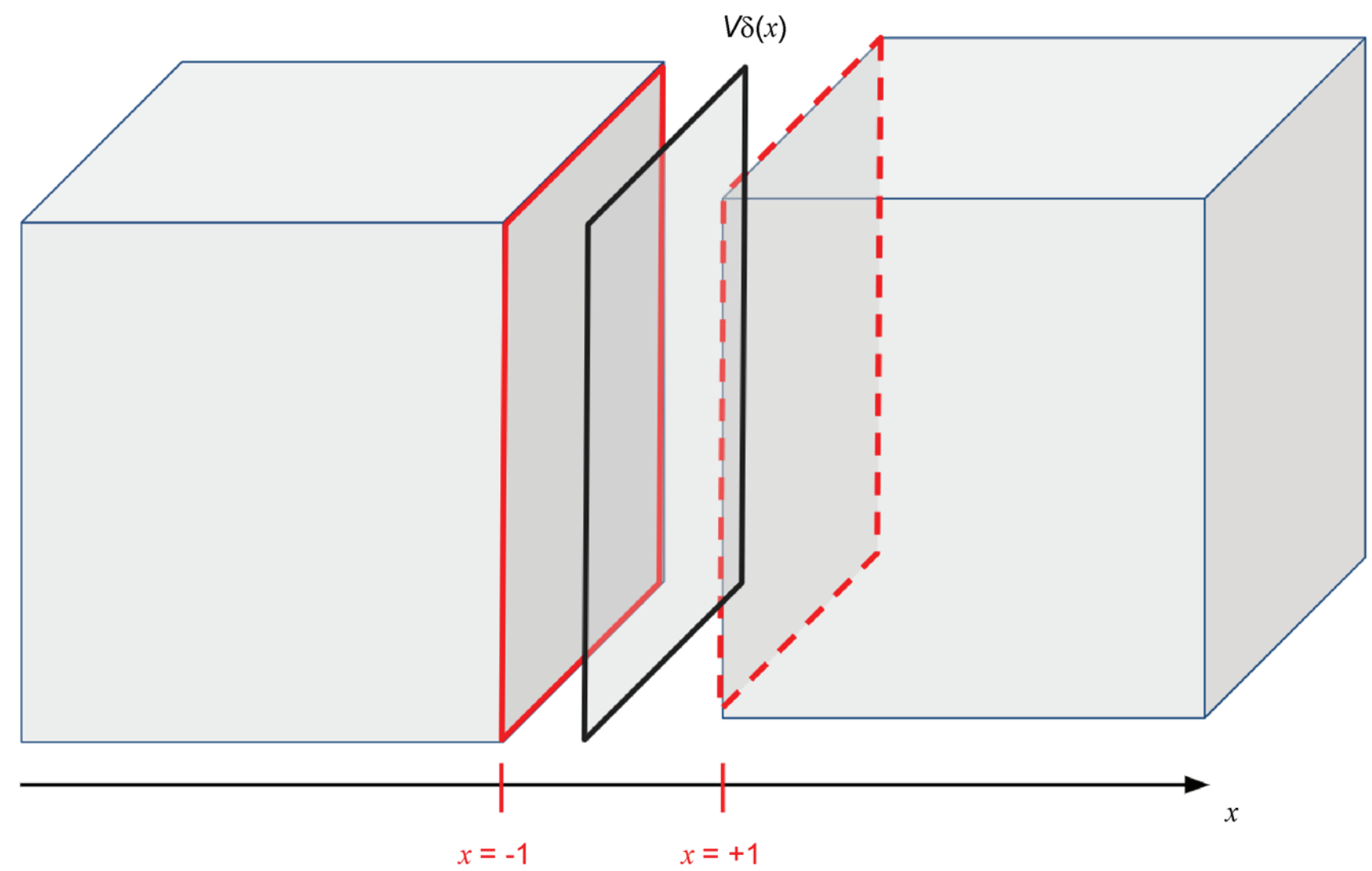

**Extended Data Fig. 2 | Schematics of the 3D system and the technique to compute the surface Green's functions.** The black parallelogram denotes the planar-impurity-potential which is oriented parallel to the (0-11) crystal plane for all calculations, while the red ones correspond to the two created surfaces on the neighboring planes at $x = \pm 1$, one lattice constant away from the impurity plane.

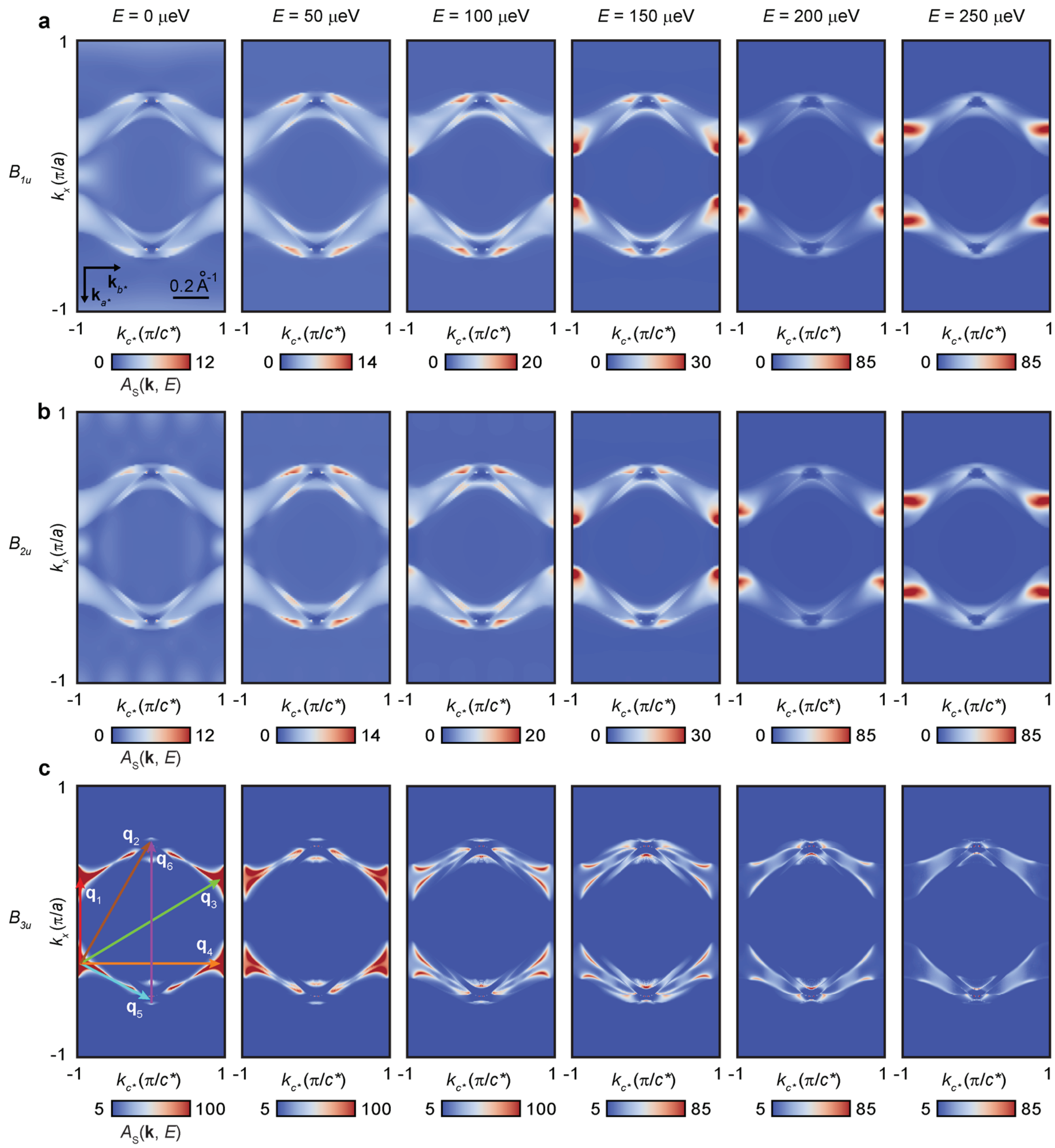

**Extended Data Fig. 3 | Surface spectral function for $B_{1u}$, $B_{2u}$ and $B_{3u}$. a.** Surface spectral function $A_s(\mathbf{k}, E)$ for $B_{1u}$ at the (0-11) SBZ. **b**. Surface spectral function $A_s(\mathbf{k}, E)$ for $B_{2u}$ at the (0-11) SBZ. **c**. Surface spectral function $A_s(\mathbf{k}, E)$ for $B_{3u}$ at the (0-11) SBZ. The anticipated sextet of scattering wavevectors $\mathbf{q}_i$, $i$ = 1-6 are overlaid.

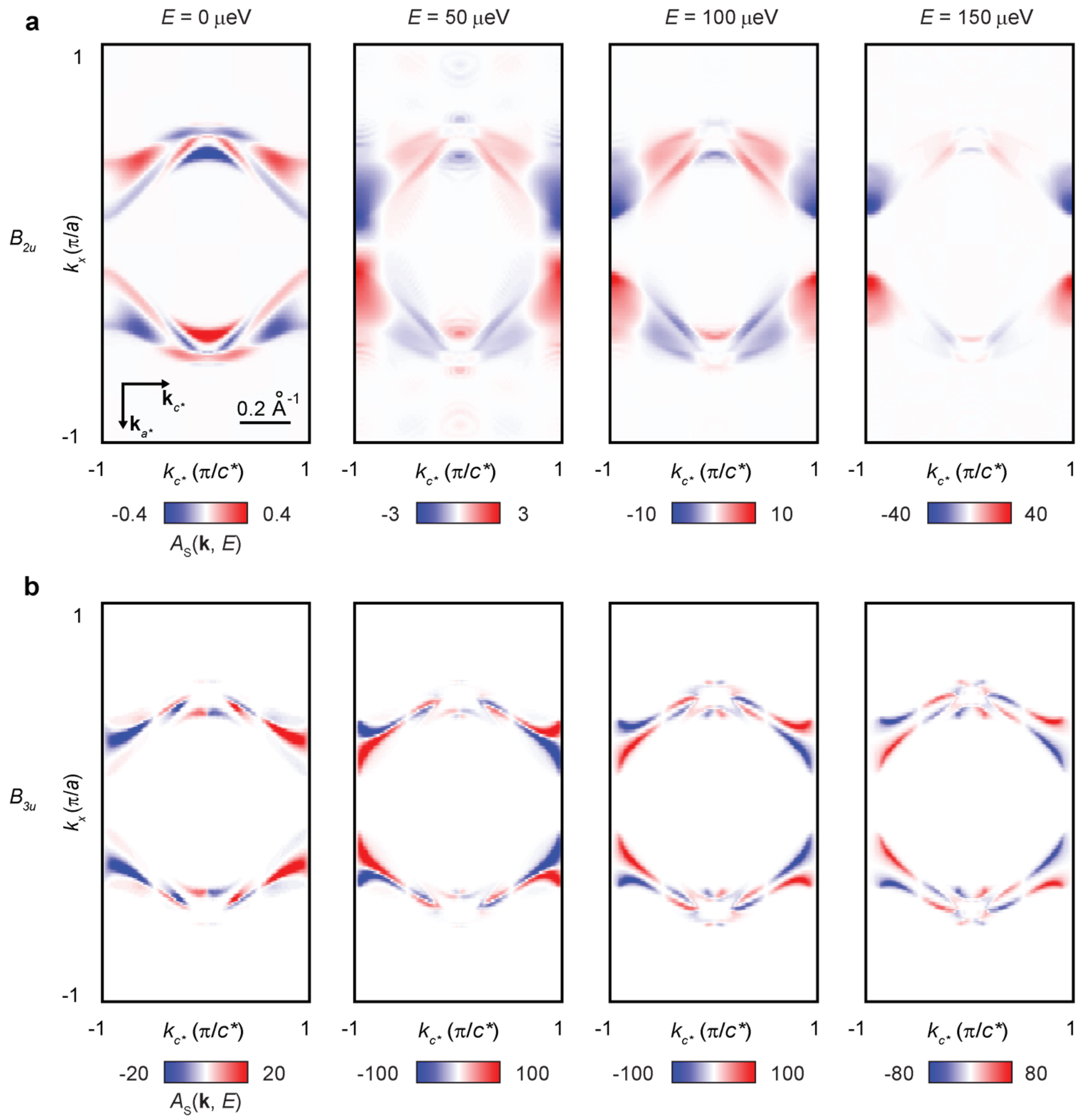

**Extended Data Fig. 4 | Spin-resolved surface spectral function for $B_{2u}$ and $B_{3u}$ gap symmetry. a**. Spin-resolved surface spectral function at the (0-11) SBZ of $UTe_2$ for a $B_{2u}$ gap function. Spin oriented parallel to the crystal **b**-axis are denoted by red and antiparallel by blue. **b**. Spin-resolved surface spectral function for a $B_{3u}$ gap function. Spin parallel to the crystal **a**-direction is denoted by red and spin antiparallel to the **a**-direction is denoted by blue. $\mathbf{q}_1$ scattering is distinct for $B_{3u}$ gap symmetry, but is not favored for $B_{2u}$ gap symmetry because spin-flip scattering processes are forbidden, as illustrated in main text Figs. 4g, i.

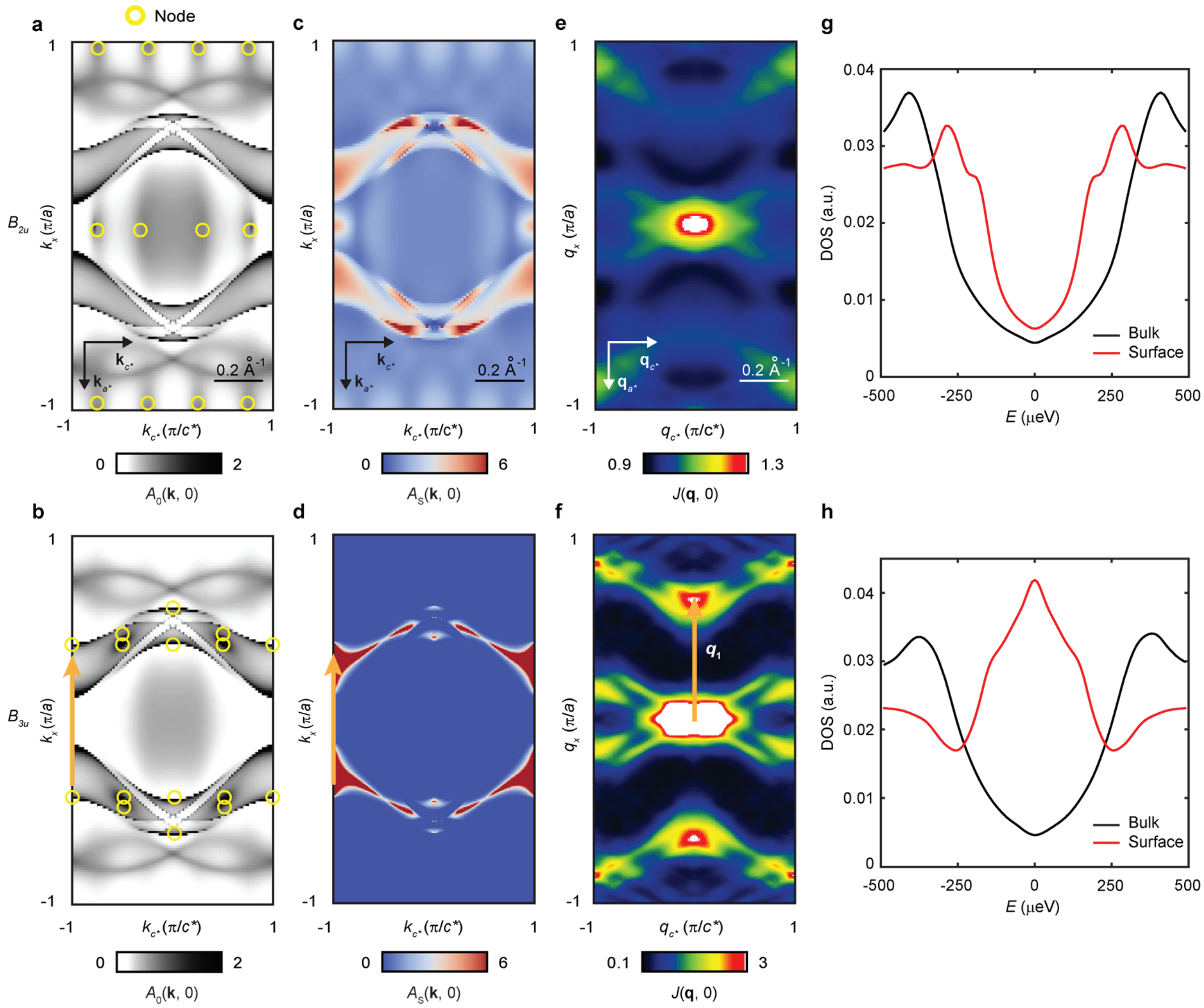

**Extended Data Fig. 5 | JDOS, QPI and DOS simulations for *B2u* and *B3u* gap symmetry. a-b.** Bulk spectral function of $UTe_2$ projection at the (0-11) SBZ of $UTe_2$ for $B_{2u}$ and $B_{3u}$ gap structures. **c-d.** Surface spectral function $A_s$(**k**, 0) at the (0-11) SBZ of $UTe_2$. **e-f.** Simulated QPI using the $J$(**q**, 0) at the (0-11) SBZ of $UTe_2$. $\mathbf{q}_1$ scattering is distinct for $B_{3u}$ gap symmetry. **g-h.** Bulk and surface band DOS calculations for (**g**) $B_{2u}$ and (**h**) $B_{3u}$ with energy gap $\Delta =$ 300 μeV. Both gap symmetries show similar bulk DOS. At the (0-11) surface, the $B_{3u}$ surface state contributes significantly to the zero-energy DOS while the $B_{2u}$ surface state is expected to have a much weaker contribution.

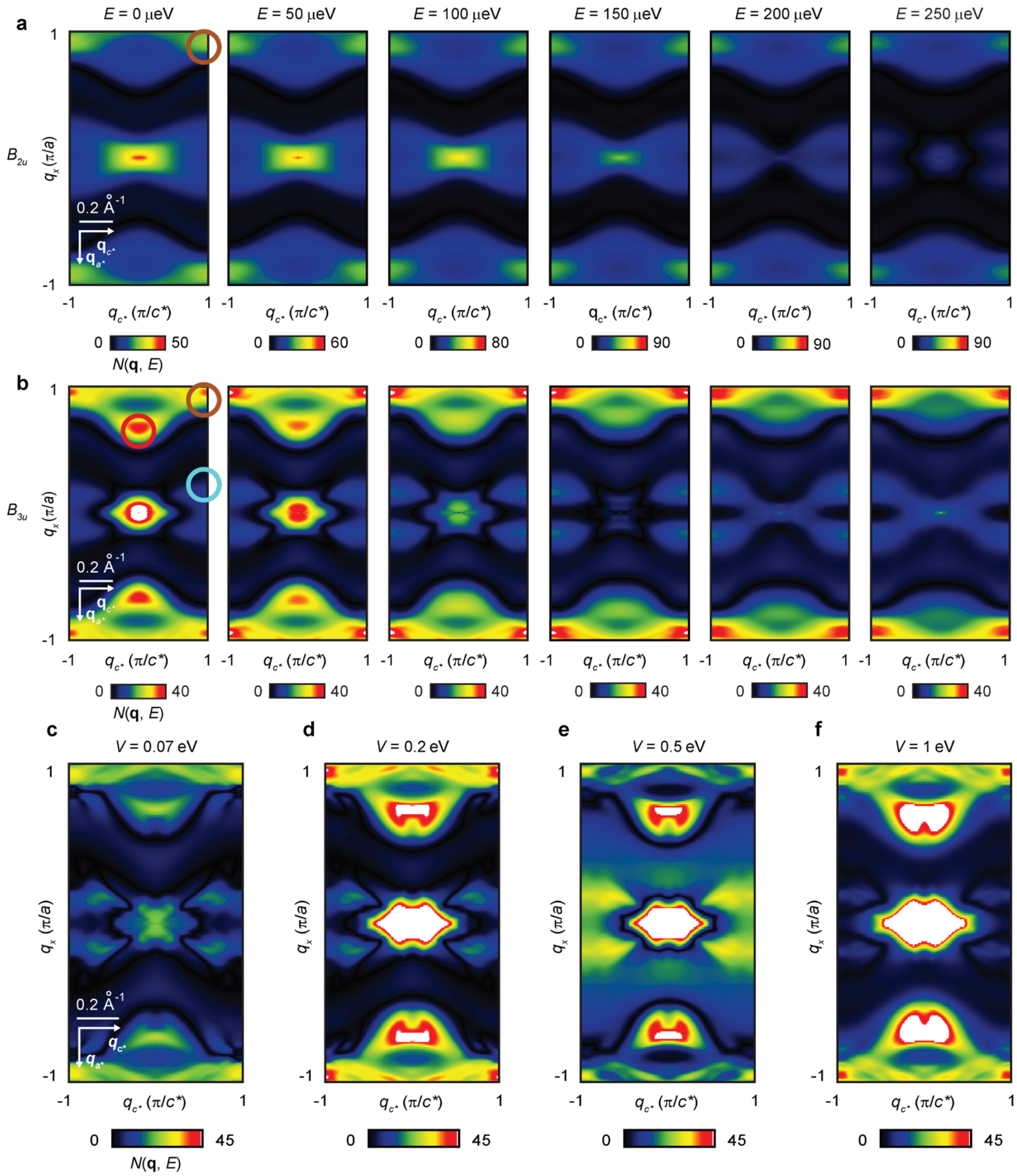

**Extended Data Fig. 6 | QPI simulations with 2D Gaussian background caused by the 'aperture' of the scan tip, and QPI simulations for *B*$_{3u}$ order parameter using different scattering potentials. a.** Unfiltered QPI simulations $N(\mathbf{q}, E)$ for a *B*$_{2u}$-QSB at the (0-11) SBZ of $UTe_2$. The existing QPI wavevector $\mathbf{q}_2$ is identified as the maxima position (brown circle). **b.** Unfiltered QPI simulations $N(\mathbf{q}, E)$ for a *B*$_{3u}$-QSB at the (0-11) SBZ of $UTe_2$. Each existing QPI wavevector, $\mathbf{q}_1$ (red), $\mathbf{q}_2$ (brown) and $\mathbf{q}_5$ (cyan), is identified as the maxima position (colored circles). To generate Figs. 4g, i of the main text, these images are multiplied by using a 2D circular Gaussian function. **c-f.** QPI simulations are performed using scattering potentials of $V$ = 0.07 eV, 0.2 eV, 0.5 eV, and 1 eV. Across this range, the predicted scattering wavevectors remain highly consistent. Varying the scattering potentials only leads to the change of the relative amplitudes of the wavevectors. The typical $N(\mathbf{q}, 50\ \mu\text{eV})$ are presented here and $\eta = 30\ \mu\text{eV}$ are used in the calculations.

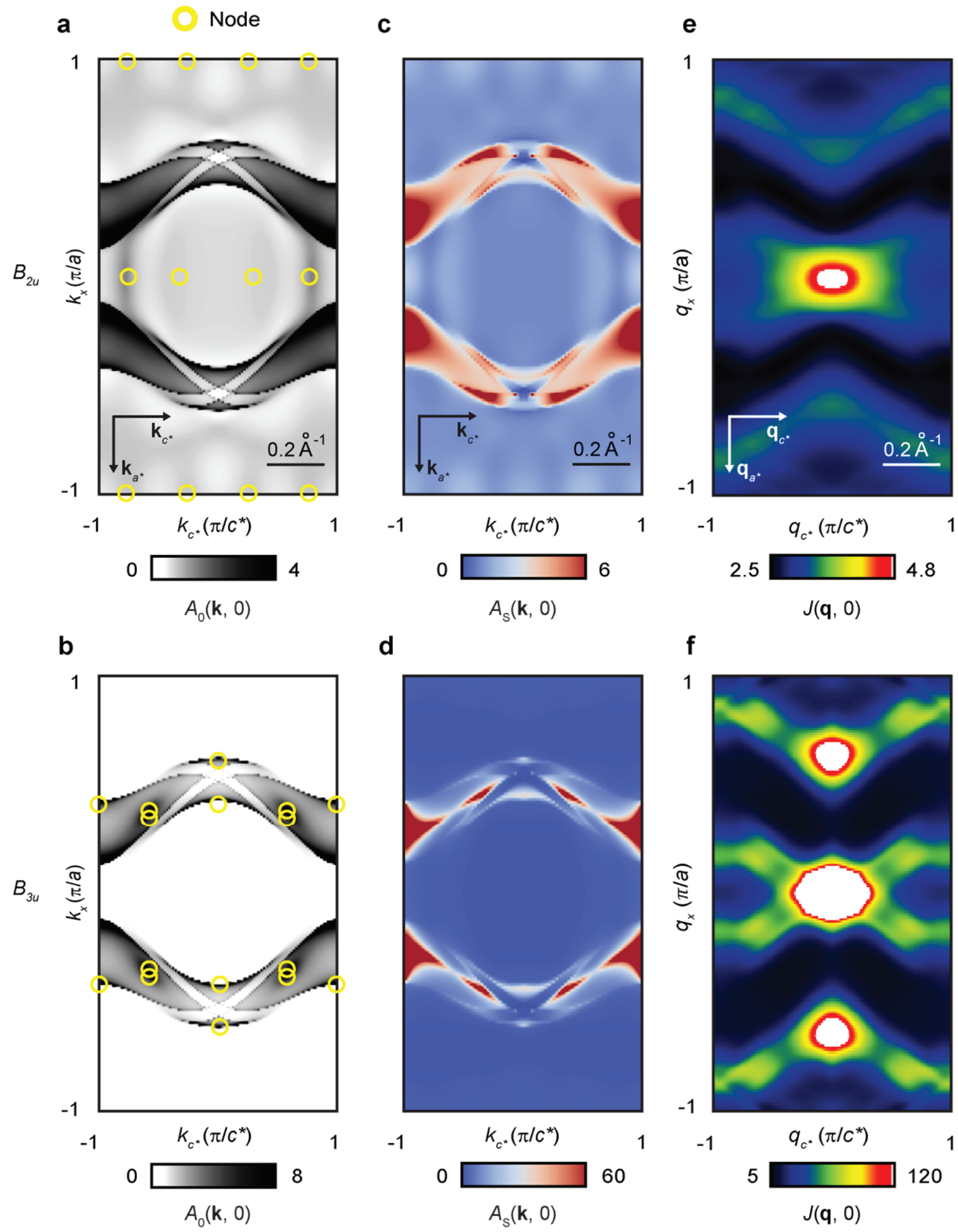

**Extended Data Fig. 7 | QPI simulations for *B2u* and *B3u* gap symmetry with alternative d-vector. a**. Bulk projected spectral function, $A_0(\mathbf{k}, 0)$ of the Fermi surface model described above with an alternative, symmetry-allowed *B2u* **d**-vector. The locations of projected nodes are highlighted with yellow circles. **b**. $A_0(\mathbf{k}, 0)$ at the (0-11) plane with an alternative **d**-vector of *B3u* symmetry. The nodal locations in **a** and **b** are very similar to those obtained using the **d**-vectors in Methods Table 1. **c**. Surface spectral function $A_s(\mathbf{k}, 0)$ for the alternative *B2u* **d**-vector. The QSB occupies regions connecting the projection of bulk nodes. **d**. $A_s(\mathbf{k}, 0)$ for the alternative *B3u* **d**-vector. The QSB develops on similar regions of the SBZ as in the Main Text. **e**. Joint density of states $J(\mathbf{q}, 0)$ of **c** for *B2u*. **f.** Joint density of states $J(\mathbf{q}, 0)$ of **d** for *B3u*. Note $\mathbf{q}_1$ scattering remains distinct for *B3u* gap symmetry.

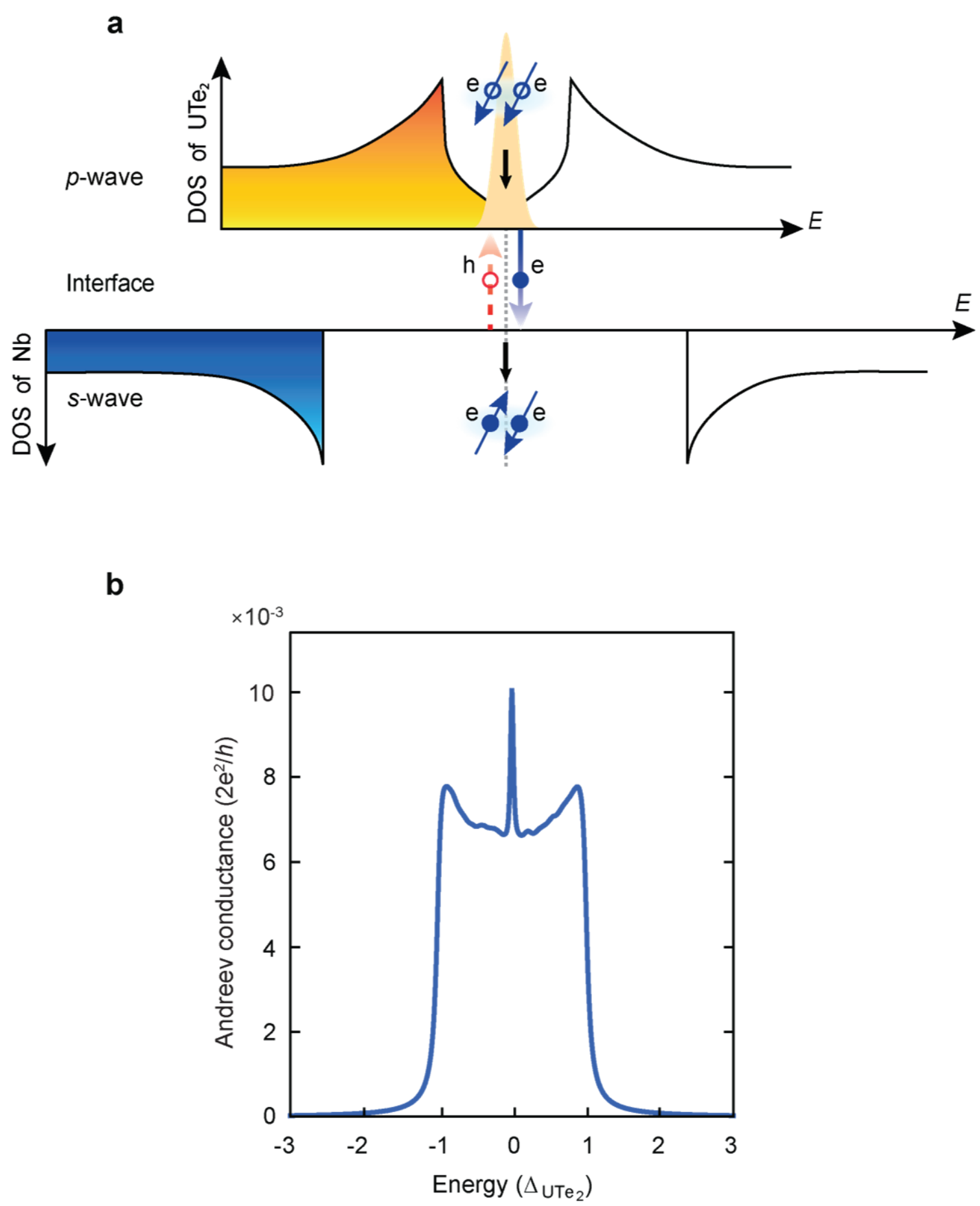

**Extended Data Fig. 8 | QSB generated Andreev conductance within SIP model. a.** Schematic of the $UTe_2$ QSB and Andreev tunneling to the *s*-wave electrode, through a two quasiparticle transport process. **b.** Calculated Andreev conductance $a(V)$ in the SIP model. The SIP model predicts a sharp peak in Andreev conductance surrounding zero-bias if the QSB is that of a *p*-wave, nodal, odd-parity superconductor that mediates the *s*-wave to *p*-wave electronic transport processes.

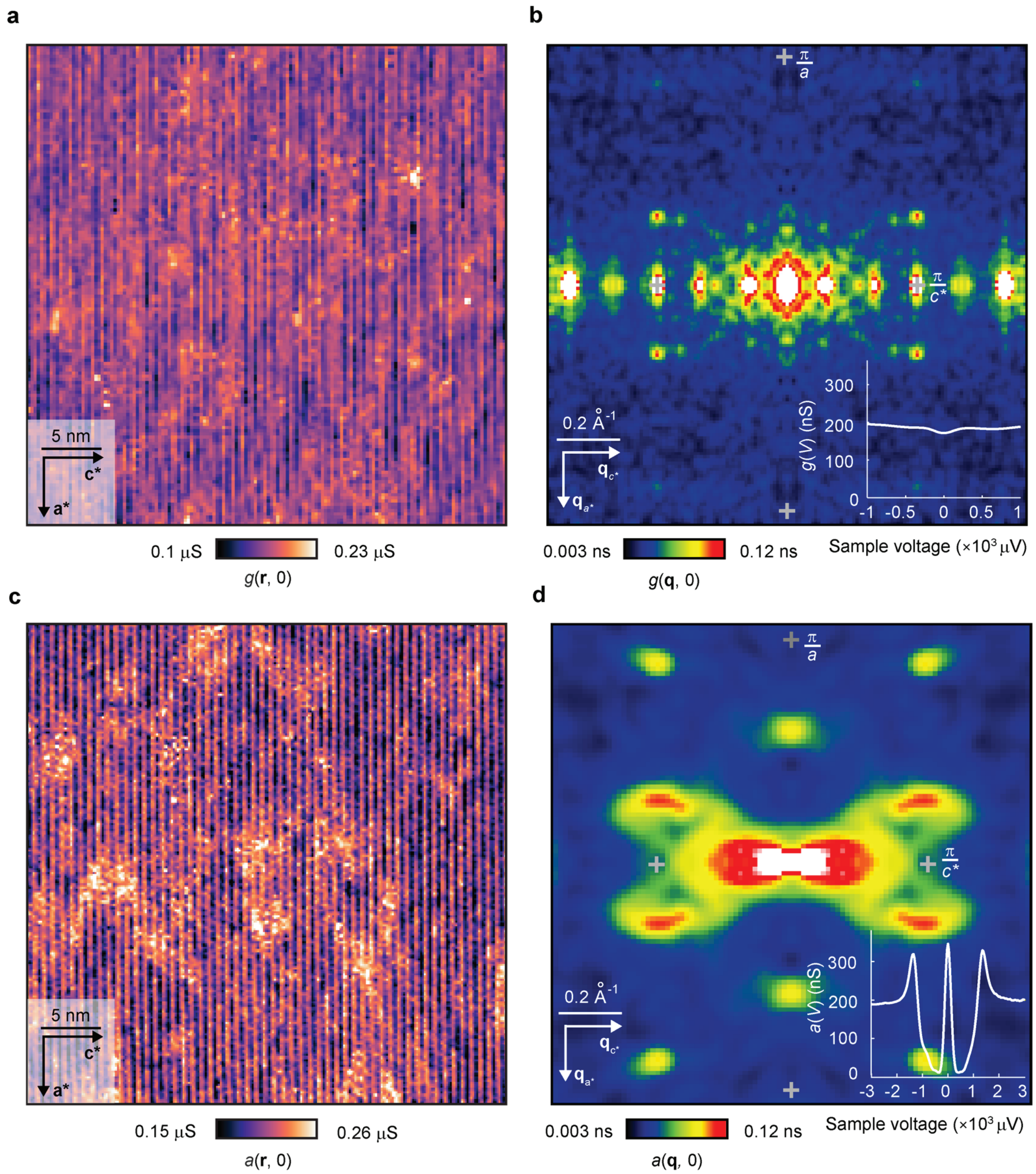

**Extended Data Fig. 9 | Non-superconductive tip and superconducting tip QSB scattering interference detection. a.** Measured normal-tip $g(\mathbf{r}, 0)$ at $T$ = 280 mK. **b.** Measured normal tip $g(\mathbf{q}, 0)$ at $T$ = 280 mK. Inset: Normal-tip single-electron tunneling spectrum $g(V)$. **c.** Measured superconductive-tip $a(\mathbf{r}, 0)$ at $T$ = 280 mK. **d.** Measured superconductive-tip $a(\mathbf{q}, 0)$ at $T$ = 280 mK. Inset: superconductive-tip Andreev tunneling spectrum $a(V)$ as described in detail in Ref. 35.

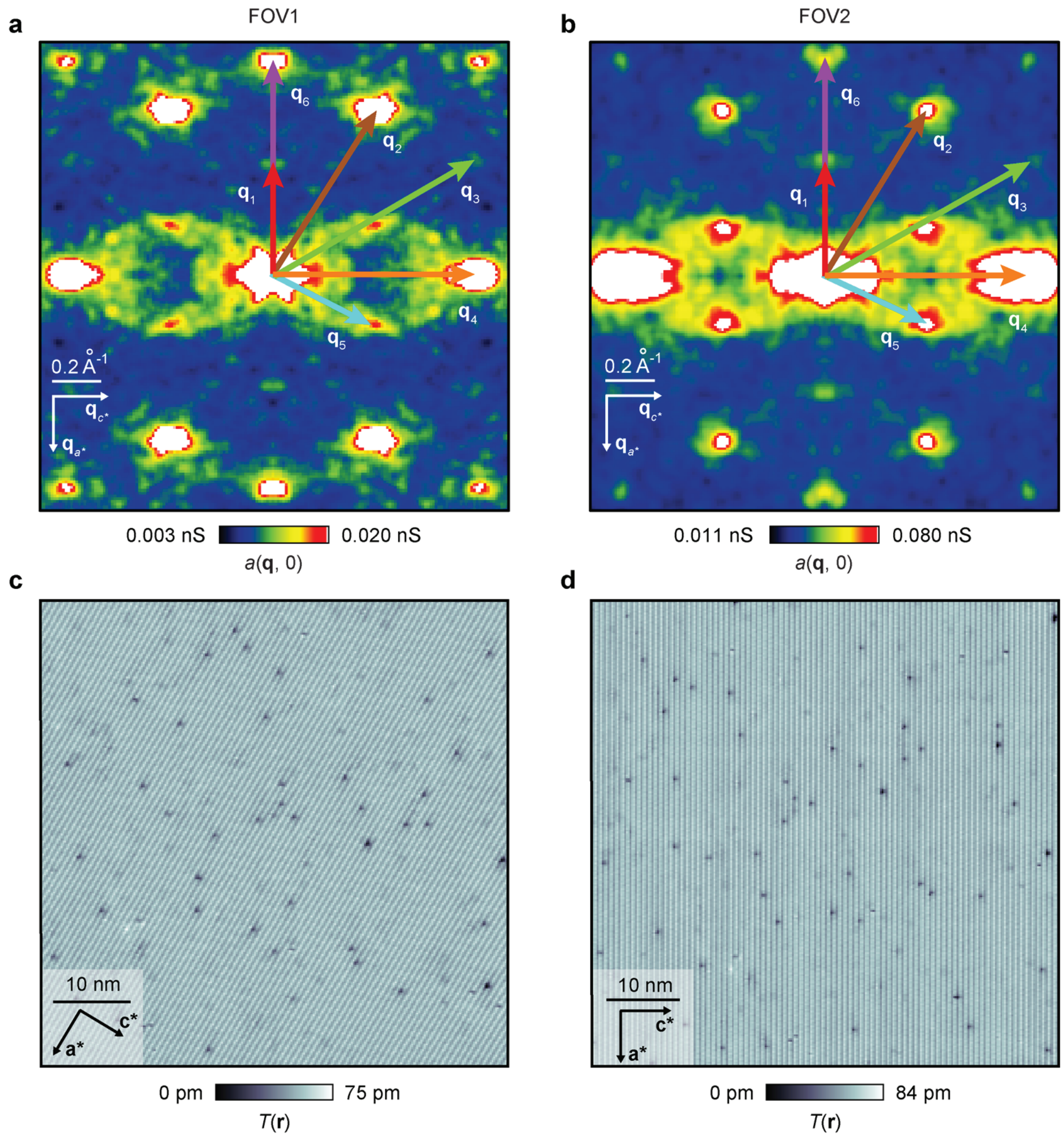

**Extended Data Fig. 10 | Independent QSB QPI visualization experiments**. (**a-b**) Two independent measurements of $a(\mathbf{q}, 0)$ at $T$ = 280 mK confirm the repeatability of the sextet of scattering wavevectors for the $B_{3u}$-QSB. (**c-d**) Topographs of areas studied in **a** and **b**, respectively.

**Data availability:** The source data shown in the main figures and Extended Data figures are available from Zenodo[63].

**Code availability:** The source code used to perform the calculations described in this paper is available from the corresponding authors upon request.

**References**

48 Theuss, F. et al. Single-Component Superconductivity in $UTe_2$ at Ambient Pressure. *Nat. Phys.* **20**, 1124–1130 (2024).

49 Peng, Y., Bao, Y., & von Oppen, F. Boundary Green functions of topological insulators and superconductors. *Phys. Rev. B* **95**, 235143 (2017).

50 McElroy K. et al., Elastic Scattering Susceptibility of the High Temperature Superconductor $Bi_2Sr_2CaCu_2O_{8+\delta}$: A Comparison between Real and Momentum Space Photoemission Spectroscopies, *Phys. Rev. Lett.* **96**, 067005 (2006).

51 Mazin, I. I., Kimber, S. A. J., Argyriou, D. N., Quasiparticle interference in antiferromagnetic parent compounds of iron-based superconductors, *Phys. Rev. B* **83**, 052501 (2011).

52 Eich, A. et al., Intra- and interband electron scattering in a hybrid topological insulator: Bismuth bilayer on $Bi_2Se_3$, *Phys. Rev. B* **90**, 155414 (2014).

53 Fang, C. et al., Theory of quasiparticle interference in mirror-symmetric two-dimensional systems and its application to surface states of topological crystalline insulators, *Phys. Rev. B* **88**, 125141 (2013).

54 Morali, N. et al., Fermi-arc diversity on surface terminations of the magnetic Weyl semimetal $Co_3Sn_2S_2$. *Science* **365**, 1286 (2019).

55 Kang, S-. H. et al., Reshaped Weyl fermionic dispersions driven by Coulomb interactions in $MoTe_2$, *Phys. Rev. B* **105**, 045143 (2022).

56 Tanaka, Y. and Kashiwaya, S., Theory of Tunneling Spectroscopy of $d$-Wave Superconductors, *Phys. Rev. Lett.* **74**, 3451 (1995).

57 Kashiwaya, S. and Tanaka, Y., Tunnelling effects on surface bound states in unconventional superconductors, *Rep. Prog. Phys*. **63**, 1641 (2000).

58 Hu, C. R., Midgap surface states as a novel signature for $d_{x_a^2-x_b^2}$-wave superconductivity, *Phys. Rev. Lett*. **72**, 1526 (1994).

59 Alff, L. et al., Spatially continuous zero-bias conductance peak on (110) $YBa_2Cu_3O_{7-\delta}$ surfaces, *Phys. Rev. B* **55**, R14757 (1997).

60 Wei, J. Y. T., et al., Directional Tunneling and Andreev Reflection on $YBa_2Cu_3O_{7-\delta}$ Single Crystals: Predominance of $d$-Wave Pairing Symmetry Verified with the Generalized Blonder, Tinkham, and Klapwijk Theory, *Phys. Rev. Lett.* **81**, 2542 (1998).

61 Liu, X. et al., Discovery of a Cooper-pair density wave state in a transition-metal dichalcogenide, *Science* **372**, 1447 (2021).

62 Agterberg, D. F. et al., The Physics of Pair Density Waves, *Annual Review of Condensed Matter Physics* **11**, 231 (2020).

63 Wang, S. et al., Data from 'Odd-Parity Quasiparticle Interference in the Superconductive Surface State of $UTe_2$'. https://doi.org/10.5281/zenodo.15597299 (2025).